\documentclass{article} 
\usepackage{latexsym,saad}

\usepackage{amssymb,amsmath,amsfonts,url} 
\usepackage[pdftex]{graphicx} 
\usepackage[ruled,vlined,linesnumbered]{algorithm2e}
\usepackage{caption}
\usepackage{verbatim} 
\usepackage{color}          
\usepackage{cite}
\usepackage{framed}

\usepackage{rotating}

\usepackage[table]{xcolor}
\usepackage{fancyhdr}

\usepackage{fancyvrb}  
   
\newcommand{\col}{\cellcolor{gray!15}}

\usepackage{wrapfig,lipsum} 

\usepackage{multirow}
\graphicspath{{./FIGS/}} 
\usepackage{graphicx,epsf,picinpar}
\usepackage[T1]{fontenc}
\usepackage{lmodern}
\usepackage{listings}
\usepackage[margin=1.0in]{geometry}
\definecolor{mygreen}{rgb}{0,0.6,0}
\definecolor{mygray}{rgb}{0.5,0.5,0.5}
\definecolor{mymauve}{rgb}{0.58,0,0.82}

\lstnewenvironment{F90}
{\lstset{
basicstyle=\footnotesize,        
  breakatwhitespace=false,         
  breaklines=false,                 
  captionpos=b,                    
  commentstyle=\color{mygreen},    
  extendedchars=true,              
  keepspaces=true,                 
  keywordstyle=\color{blue},       
  language=[95]Fortran,                 
  numbers=left,                    
  numbersep=5pt,                   
  numberstyle=\tiny\color{mygray}, 
  rulecolor=\color{black},         
  showspaces=false,                
  showstringspaces=false,          
  showtabs=false,                  
  stepnumber=1,                    
  stringstyle=\color{mymauve},     
  tabsize=4,                       
  title=\lstname                   
}}
{}
\lstnewenvironment{C}
{\lstset{language=C++,
                basicstyle=\footnotesize,
                keywordstyle=\color{blue},
                stringstyle=\color{mymauve},
                commentstyle=\color{mygreen},
                  numbers=left,                    
  numbersep=5pt,                   
  numberstyle=\tiny\color{mygray}, 
                morecomment=[l][\color{magenta}]{\#}
}}
  {}

\colorlet{shadecolor}{blue!10}

\usepackage{framed}

\newsavebox{\fmbox}

\newcommand{\versionT}{{4.0}}
\newcommand{\version}{{v\versionT}{\ }}
\newcommand{\feastdir}{{\tt <}{\it {\tt FEAST} directory}{\tt >~}}
\newcommand{\arch}{{\tt <{\it arch}>~} }

\begin{document}

\thispagestyle{empty}

\centerline{\sc High-Performance Numerical Library for Solving Eigenvalue Problems}

\vspace{4cm}
\begin{center}
{\Huge FEAST Eigenvalue Solver \version\\[10pt]
User Guide} \\
\vspace{2.2cm}
\scalebox{0.7}{\includegraphics{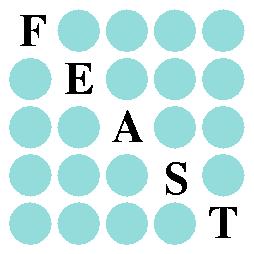}}\\
{\tt http:://www.feast-solver.org}
\end{center}

\vspace{3.5cm}
\begin{center}
\begin{Large}
Eric Polizzi's Research Lab. \\
Department of Electrical and Computer Engineering, \\
Department of Mathematics and Statistics, \\[5pt]
University of Massachusetts, Amherst
\end{Large}
\end{center}

\newpage

\section*{References}

\noindent If you are using FEAST, please consider citing one or more publications below in your work.

\begin{small}
\noindent \begin{description}
\itemsep 1pt
\parskip 1pt
\item[{\bf Main reference}]~ \\ E. Polizzi, {\it Density-Matrix-Based Algorithms for Solving Eigenvalue Problems},\\
 Phys. Rev. B. Vol. 79, 115112 (2009)
\item[{\bf Math analysis}]~ \\ P. Tang, E. Polizzi, {\it FEAST as a Subspace Iteration EigenSolver Accelerated by Approximate Spectral Projection};
SIAM Journal on Matrix Analysis and Applications (SIMAX)  35(2), 354-390 - (2014) 

\item[{\bf Non-Hermitian solver}]~ \\ J. Kestyn, E. Polizzi, P. T. P. Tang, 
{\it FEAST Eigensolver for Non-Hermitian Problems}, \\ 
SIAM Journal on Scientific Computing (SISC), 38-5, ppS772-S799
(2016);

\item[{\bf Hermitian using Zolotarev quadrature}]~ \\ S. G\"uttel, E. Polizzi, P. T. P. Tang, G. Viaud, {\it
  Optimized Quadrature Rules and Load Balancing for the FEAST Eigenvalue Solver}, \\
  SIAM Journal on Scientific Computing (SISC), 37 (4),
pp2100-2122 (2015).

\item[{\bf Eigenvalue count using stochastic estimates}]~ \\ 
E. Di Napoli, E. Polizzi, Y. Saad, {\it
Efficient Estimation of Eigenvalue Counts in an Interval},\\ 
Numerical Linear Algebra with Applications, V23, I4, pp674-692,(2016).

\item[{\bf Polynomial Non-linear eigenvalue problem -- Residual Inverse Iterations}]~ \\
  B. Gavin, A. Miedlar, E. Polizzi,{\it FEAST Eigensolver for Nonlinear Eigenvalue Problems
    }\\
  Journal of Computational Science, V. 27, 107,
  (2018)

  \item[{\bf IFEAST}]~ \\
  B. Gavin, E. Polizzi,{\it Krylov eigenvalue strategy using the FEAST algorithm with inexact system solves}\\
  Numerical Linear Algebra with Applications, vol 25, number 5, 20 pages (2018).
  
 \item[{\bf PFEAST}]~ \\
   J. Kesyn, V. Kalantzis, E. Polizzi, Y. Saad,{\it PFEAST: A High Performance Sparse Eigenvalue Solver Using Distributed-Memory Linear Solvers}\\
   Proceedings of the International Conference for High
Performance Computing, Networking, Storage and Analysis, ACM/IEEE Supercomputing Conference
(SC’16), pp 16:1-16:12, (2016).


\end{description}
\end{small}


\section*{Contact}

If you have any questions or feedback regarding FEAST, please send an-email to  
{\bf feastsolver@gmail.com}.

\vspace{0.5cm}

\section*{FEAST algorithm and software team, collaborators and contributors}

\begin{description}
\itemsep 1pt
\parskip 1pt
\item[Code Developer/Contributors] \begin{minipage}[t]{0.65\linewidth}
  Eric Polizzi (Lead) \\
  James Kestyn (Non-Hermitian, PFEAST),\\
  Brendan Gavin (Non-linear, IFEAST), \\
  Braegan Spring (SPIKE banded solver, Hybrid solver--in progress),\\
  Stefan G\"uttel (Zolotarev quadrature),\\
  Julien Brenneck (GUI configurator,  Quaternion--in progress).
  \end{minipage}
\item[Collaborators]: Peter Tang, Yousef Saad, Agnieszka Miedlar, Edoardo Di Napoli, Ahmed Sameh
 \end{description}

\vspace{1cm}

\section*{Acknowledgments}


This work has been supported by Intel Corporation and by the National Science Foundation (NSF) 
under Grants \#CCF-1510010, \#SI2-SSE-1739423, and \#CCF-1813480.

\newpage

\begin{Large}
\tableofcontents
\end{Large}
\newpage



\section{Background}

{\small {\em 
``The solution of the algebraic eigenvalue problem has for long had a
particular fascination for me because it illustrates so well the difference
between what might be termed classical mathematics and practical
numerical analysis. The eigenvalue problem has a deceptively simple
formulation and the background theory has been known for many
years; yet the determination of accurate solutions presents a wide
variety of challenging problems.''}\\
J. H. Wilkinson- The Algebraic Eigenvalue Problem- 1965}

\vspace{0.3cm}

The eigenvalue problem is ubiquitous in  science and engineering applications. It can be encountered under different forms:
Hermitian or non-Hermitian, and linear or non-linear. The eigenvalue problem has led to many challenging numerical questions and a
central problem: how can we compute eigenvalues and eigenvectors in an efficient manner and how
accurate are they? 

The FEAST library package represents an unified framework for solving various family of eigenvalue problems and
addressing the issues of numerical accuracy, robustness, performance and parallel scalability.
 Its  originality lies with a new transformative numerical approach to the
traditional eigenvalue algorithm design - the FEAST algorithm.

\subsection{The FEAST Algorithm}

The FEAST algorithm is a general purpose eigenvalue solver
 which takes its inspiration from the density-matrix
 representation and contour integration technique in quantum mechanics\footnote{E. Polizzi, Phys. Rev. B. Vol.  79, 115112 (2009)}.
 The algorithm gathers key elements
from complex analysis, numerical linear algebra and approximation theory, to construct an optimal subspace
iteration technique making use of approximate spectral projectors\footnote{P. Tang, E. Polizzi, SIMAX  35(2), 354–390 - (2014)}.
FEAST can be applied for solving  both standard and generalized forms of the Hermitian or non-Hermitian problems (linear or non-linear),
 and it belongs to the family of contour integration eigensolvers.  
Once a given search interval is selected, FEAST's main computational task consists of a numerical 
quadrature computation that involves solving independent linear systems along a complex contour, each with
multiple right hand sides. 
A Rayleigh-Ritz procedure is then used to generate
 a reduced dense eigenvalue problem orders of magnitude smaller 
than the original one
 (the size of this reduced problem is of the order of the
 number of eigenpairs inside the search interval/contour).
FEAST offers a set of appealing features: (i)
Remarkable robustness with well-defined convergence
rate; (ii) All multiplicities naturally captured;
(iii) No explicit orthogonalization procedure on long vectors
required in practice; (iv) Reusable subspace as initial guess when solving a series of eigenvalue problems; and (v)
Efficient use of both blocked BLAS-3 operations and parallel resources for solving the
linear systems with multiple right hand sides.
FEAST can exploit   a key strength of modern computer architectures,
namely, multiple levels of parallelism. Natural parallelism appears at three different levels (L1,
L2 or L3): (L1) search contours can be treated separately
(no overlap), (L2) linear systems can be solved independently
across the quadrature nodes of the complex contour, and (L3)
each complex linear system with multiple right-hand-sides can
be solved in parallel. Parallel resources can be placed at all
three levels simultaneously in order to achieve scalability and
optimal use of the computing platform.  Within a parallel environment, the main numerical task can be
 reduced to the solution of a single linear system 
 using direct or iterative parallel solvers.


\subsection{The FEAST Solver}

FEAST release dates with main features are listed below:
\begin{enumerate}
\itemsep 1pt
\parskip 1pt
\item[\bf v1.0] (Sep. 2009): Hermitian problem (standard/generalized)
\item[\bf v2.0] (Mar. 2012): SMP+MPI+RCI interfaces
\item[\bf v2.1] (Feb. 2013): {\bf Adoption by Intel-MKL}
\item[\bf v3.0] (Jun. 2015): Support for non-Hermitian
\item[\bf v4.0] (Feb. 2020): Residual inverse iterations - mixed precision - IFEAST (FEAST w/o factorization) - PFEAST (3 MPI levels) -
  Support  for non-linear (polynomial) - Support for extreme eigenvalues (lowest/largest)
\end{enumerate}

\newpage

The FEAST package v2.1 has been featured as Intel-MKL's principal HPC eigensolver since  2013\footnote{https://software.intel.com/en-us/articles/introduction-to-the-intel-mkl-extended-eigensolver}.  The current version of the FEAST package (v4.0) released in Feb. 2020
represents a significant upgrade since the entire FEAST package has been re-coded to perform residual
inverse iterations\footnote{B. Gavin, A. Miedlar, E. Polizzi, Journal of Computational Science, V. 27, 107,
  (2018); B. Gavin, E. Polizzi, in preparation (2020).}. As a result, v4.0 is in average much faster
than v2.1 or v3.0 ($\times 3-4$ using new default optimization parameters), and it became possible to implement new
important features such as IFEAST (using Inexact Iterative solver) and Non-linear polynomial FEAST. Furthermore, v4.0 features PFEAST
with its 3-MPI levels of parallelism.

\begin{figure}[htbp]
\begin{minipage}{0.53\linewidth}
\centering
\includegraphics[width=0.9\linewidth]{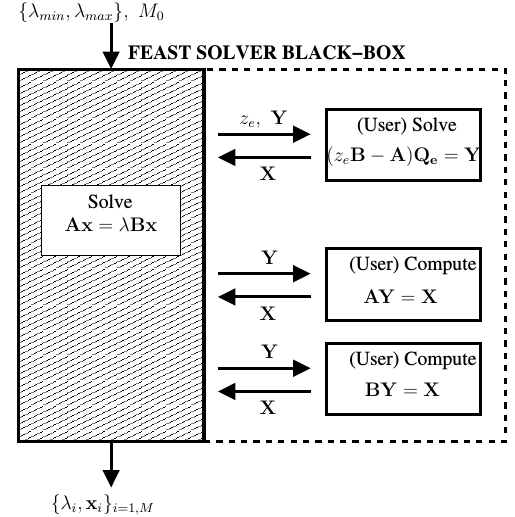}
\end{minipage}
\hfill
\begin{minipage}{0.46\linewidth}
\caption{\label{fig_bb} \small \em ``Black-box'' interface for the Hermitian problem. In normal mode,
FEAST requires a search interval and a search subspace size $M_0$. 
It includes features such as reverse communication interfaces (RCI)
that are matrix format independent, and linear
system solver independent, as well as ready to
use driver interfaces for dense, banded and
sparse systems. For the driver interfaces
the ``black-box'' region extends then to the right
dashed box, and only the system matrices are 
required as inputs from the users. The RCI 
interfaces represent the kernel of FEAST which can be
customized by the users to allow maximum 
flexibility for their specific applications. Users have
then the possibility to integrate their own linear
system solvers (direct or iterative - with or without preconditioner) and handle their own matrix-vector multiplication procedure.}
\end{minipage}
\end{figure}

FEAST is a comprehensive numerical library offering both simplicity and flexibility, and packaged around a
``black-box'' interface as depicted in Figure \ref{fig_bb} for the Hermitian problem.

The current main features of the FEAST v4.0 package include: 
\begin{itemize}
\itemsep 1pt
\parskip 1pt 
\item Standard or Generalized Hermitian and non-Hermitian eigenvalue problems (left/right eigenvectors and bi-orthonormal basis);
\item Polynomial eigenvalue problems such as quadratic, cubic, quartic, etc. (left/right eigenvectors);
\item Finding eigenpair within a search contour (normal mode); Finding extreme eigenvalues (lowest/largest) for sparse Hermitian systems;  
\item Real/Complex arithmetic and mixed precision (single precision operations leading to double precision final results);
\item Two libraries: {\bf SMP version} (one node), and  {\bf MPI version} (multi-nodes);  
\item Reverse communication interfaces (RCI). 
\item Driver interfaces for dense (using LAPACK), banded (using SPIKE), and sparse-CSR formats (using MKL-PARDISO);  
\item IFEAST- FEAST w/o factorization for sparse-CSR drivers (using BiCGStab); 
\item PFEAST- FEAST using 3 levels of MPI parallelism for HPC (MPI solvers includes MKL-Cluster-PARDISO and PBiCGStab); Sparse and RCI interfaces compatible with local row-distributed data.
\item A set of flexible and useful practical options (quadrature rules, contour shapes,  stopping criteria, initial guess, fast stochastic estimates for eigenvalue counts, etc.) 
\item Portability: FEAST routines can be called from any Fortran or C codes.
\item FEAST interfaces only require (any optimized) LAPACK and BLAS packages.
\item Large number of driver examples, utility routines, and documentation.
\end{itemize}




\newpage 

\section{Installation and Setup: A Step by Step Procedure}


\subsection{Installation}

Please follow the following steps (here for Linux/Unix systems):                 
\begin{enumerate}
\item Download the latest FEAST package version {\tt feast\_4.0.tgz} in {\bf http://www.feast-solver.org}
\item Put the file in your preferred directory such as {\tt \$HOME} directory or (for example)  {\tt /opt/} directory
 if you have ROOT privilege. 
\item Execute: {\tt \bf tar -xzvf feast\_4.0.tgz}  
to create the following {\tt FEAST} tree directory.

\begin{shaded}
\begin{center}
    \begin{verbatim}
                                FEAST
                                  |
                                 4.0 
                                  |
     --------------------------------------------------------------
     |         |           |             |           |            |
    doc     example     include         lib         src        utility
               |                         |           |            |
         -------                  --------     -------        -----                 
        |-FEAST                  |-x64        |-kernel       |-FEAST        
        |-PFEAST-L2                           |-dense        |-PFEAST
        |-PFEAST-L2L3                         |-banded       |-data
        |-PFEAST-L1L2L3                       |-sparse
\end{verbatim}        
\end{center}
\end{shaded}

\item If \feastdir  denotes the package's main directory after
installation, for example
\begin{quote}
{\tt $\sim$/home/FEAST/4.0}~~~ or ~~~{\tt /opt/FEAST/4.0},
\end{quote}
{\bf it is not mandatory but recommended to define the Shell variable {\tt \$FEASTROOT},} e.g.  
\begin{quote}
\fbox{\tt export FEASTROOT=\feastdir}~~~~or~~~~\fbox {\tt set FEASTROOT=\feastdir} 
\end{quote}
respectively  for the BASH or CSH shells.
One of this command can be placed in the appropriate shell startup file in {\tt \$HOME} (i.e {\tt .bashrc}
 or {\tt .cshrc}).


\end{enumerate}

\subsection{Compilation}

Go to the directory {\tt \$FEASTROOT/src} and execute {\tt make} to see all available options. The same FEAST  source code is used for compiling FEAST-SMP ({\tt libfeast}) and/or FEAST-MPI ({\tt libpfeast}). The command is:\\
  
  \fbox{{\tt  make ARCH=<arch> F90=<f90> MPI=<mpi> MKL=<mkl> \{feast, pfeast\}}}\\

 where you can select the following options: 
\begin{enumerate}
\itemsep 1pt
\parskip 1pt
\item[{\bf <arch>}]: it is the name of the directory {\tt \$FEASTROOT/lib/\arch} where the FEAST libraries will be located once compiled (you can use the name of
  your  architecture). Default is {\bf x64}
\item[{\bf <f90>}]: it is your own Fortran90 compiler (possible choices: ifort, gfortran, pgf90). Default is {\bf ifort}
\item[{\bf <mpi>}]: (mandatory for compiling {\tt libpfeast} only) it is your MPI library (possible choices: impi, mpich, openmpi). Defaults to {\bf impi} (intel MPI)
\item[{\bf <mkl>}]: it enables Intel-MKL math library instructions (possible choices: yes, no). Default is {\bf yes}.
  \begin{itemize}
          \item if <mkl>=yes, at the linking stage, FEAST will have to be linked with Intel MKL.
          \item if <mkl>=no, at the linking stage, FEAST can be linked with any BLAS/LAPACK. Not using MKL will
            impact the behavior and performance of the FEAST sparse Driver interfaces:
            (i) it would not be possible to use MKL-PARDISO and cluster-MKL-PARDISO so the FEAST sparse interfaces will
            instead be calling IFEAST (BiCGStab);
            (ii) in-built sparse mat-vec routines (used by the IFEAST sparse interfaces) will be slower.
\end{itemize}
\end{enumerate}

For example, if the above default options look fine with you, just use:
\begin{itemize}
\item {\tt make feast}  \\ to compile the FEAST-SMP library and create the file {\tt libfeast.a} in {\tt \$FEASTROOT/lib/\arch} 
\item {\tt make pfeast} \\ to compile the FEAST-MPI library and create the file {\tt libpfeast.a} in {\tt \$FEASTROOT/lib/\arch} 
\end{itemize}

 Congratulations, FEAST is now successfully installed and compiled  on your computer !!

\subsection{Linking FEAST}

In order to use the FEAST library for your main application code, you will then need to add the following instructions
in your {\tt Makefile}:
\begin{itemize}
  \itemsep 1pt
\parskip 1pt
\item {\it for the LIBRARY PATH:}~~ {\tt -L\$FEASTROOT/lib/\arch } 
\item {\it for the LIBRARY LINKS using FEAST-SMP:}  {\tt -lfeast}\\
{\color{white}{aaaaaaaaaaaaaaaaaaaaaaaa}}{\it using FEAST-MPI:} {\tt -lpfeast}
\item {\it for the INCLUDE PATH  (mandatory only for {\tt C} codes):}~~  {\tt  -I\$(FEASTROOT)/include} 

\end{itemize}

\noindent {\bf Remarks} \\
\noindent 1- If FEAST was compiled with the option {\tt MKL=yes}, your must also link with the MKL libraries. Otherwise,
you can link with any BLAS, LAPACK libraries.\\
\noindent 2- If you use the FEAST banded interfaces, you need to install the SPIKE solver \url{www.spike-solver.org}  SPIKE must be compiled using the same Fortran compiler used for compiling FEAST.\\
\noindent 3- For C codes, the user must include the following instructions in the header:

\begin{small}
\begin{C}
#include "feast.h"        
#include "feast_dense.h"  //for feast dense interfaces
#include "feast_banded.h" //for feast banded interfaces
#include "feast_sparse.h" //for feast sparse interfaces
\end{C}
\end{small}

\noindent Using PFEAST (MPI sparse linear system solver), you must use instead:

\begin{small}
\begin{C}
#include "pfeast.h"        
#include "pfeast_sparse.h" 
\end{C}
\end{small}

\newpage

\subsection{HelloWorld Example ({\tt F90, C, MPI-F90, MPI-C})}\label{sec-hello}

This example solves a 4-by-4 real symmetric standard eigenvalue system $\bf Ax=\lambda x$ (using dense format) where

\begin{equation}
{\bf A}=
\begin{pmatrix}
 2 & -1 & -1 & 0\\
 -1 & 3 & -1 &-1\\
 -1 &-1&3 &-1 \\
 0 &-1 & -1 & 2
 \end{pmatrix}.
\end{equation}
The four eigenvalue solutions are $\lambda=\{0,2,4,4\}$.
Let us suppose that one can specify a search interval (such as $[3,5]$), a single call to the {\tt dfeast\_syev} subroutine should 
return the solutions associated with $\{4,4\}$. The FEAST parameters need first to be set to their default values by 
a call to the {\tt feastinit} subroutine. Below, we provide examples written in {\tt F90}, {\tt C}, {\tt MPI-F90}, and {\tt MPI-C}.

\subsubsection*{F90}
A Fortran90 source code of {\tt helloworld.f90} is provided below: 

\begin{small}
  \begin{F90}%[frame=trBL]
program helloworld
  implicit none
  !! 4x4 eigenvalue system 
  integer,parameter :: N=4
  character(len=1) :: UPLO='F' ! 'L' or 'U' also fine
  double precision,dimension(N*N) :: A=(/  2.0d0, -1.0d0,-1.0d0, 0.0d0,&
                                         &-1.0d0,  3.0d0,-1.0d0,-1.0d0,&
                                         &-1.0d0, -1.0d0, 3.0d0,-1.0d0,&
                                         & 0.0d0, -1.0d0,-1.0d0, 2.0d0/)
  !! input parameters for FEAST
  integer,dimension(64) :: fpm
  integer :: M0=3 ! search subspace dimension
  double precision :: Emin=3.0d0, Emax=5.0d0 ! search interval
  !! output variables for FEAST
  double precision,dimension(:),allocatable :: E, res
  double precision,dimension(:,:),allocatable :: X
  double precision :: epsout
  integer :: loop,info,M,i

!!! Allocate memory for eigenvalues.eigenvectors,residual
  allocate(E(M0),X(N,M0),res(M0))

!!!!!!!!!! FEAST
  call feastinit(fpm)
  fpm(1)=1 !! change from default value (print info on screen)
  call dfeast_syev(UPLO,N,A,N,fpm,epsout,loop,Emin,Emax,M0,E,X,M,res,info)

!!!!!!!!!! REPORT
  if (info==0) then
     print *,'Solutions (Eigenvalues/Eigenvectors/Residuals)'
     do i=1,M
        print *,'E=',E(i),'X=',X(:,i),'Res=',res(i)
        print *,''
     enddo
  endif

end program helloworld
\end{F90}
\end{small}


\noindent To create the executable, compile and link the source program with the
{\tt feast} library, one can use (for example):
\begin{itemize}
\item \fbox{\tt ifort -o helloworld helloworld.f90 -L\$FEASTROOT/lib/<{\it arch}> -lfeast -mkl} \\
  if FEAST was compiled with {\tt ifort} and MKL flag was set to 'yes'.
\item \fbox{\parbox{\dimexpr\linewidth-2\fboxsep-2\fboxrule\relax}{\tt gfortran -o helloworld helloworld.f90 -L\$FEASTROOT>/lib/<{\it arch}> -lfeast  -Wl,--start-group -lmkl\_gf\_lp64 -lmkl\_gnu\_thread -lmkl\_core -Wl,--end-group -lgomp -lpthread -lm -ldl
}} \\
    if FEAST was compiled with {\tt gfortran} and MKL flag was set to 'yes'.
\end{itemize}  

\noindent {\bf Remarks:} \\
\noindent 1-Many other options are possible. For example, you can link with your own BLAS/LAPACK libraries, you can compile FEAST with {\tt ifort} and compile the helloworld example using {\tt gfortran} via extra flag options, or vice-versa, etc.\\
\noindent 2- FEAST is using a linear solver that can be threaded (the LAPACK dense solver is used for the helloworld example). Using MKL,
you can control the number of threads by setting up the value of {\tt MKL\_NUM\_THREADS}. For example, in BASH shell: \\

\fbox{\tt export MKL\_NUM\_THREADS=<omp>} \\

where ${\tt <omp>}$ represents the number of threads (cores). \\

\noindent A run of the resulting executable looks like:
         \begin{center}\fbox{\tt ./helloworld} \end{center}
         and the output of the run should be: 

\begin{shaded}
\begin{footnotesize}
\begin{Verbatim}[frame=single,baselinestretch=0.9]
***********************************************
*********** FEAST v4.0 BEGIN ******************
***********************************************
Routine DFEAST_SYEV
Solving AX=eX with A real symmetric
List of input parameters fpm(1:64)-- if different from default
   fpm( 1)=   1
 
.-------------------.
| FEAST data        |
--------------------.-------------------------.
| Emin              |  3.0000000000000000E+00 |
| Emax              |  5.0000000000000000E+00 |
| #Contour nodes    |  8   (half-contour)     |
| Quadrature rule   |  Gauss                  |
| Ellipse ratio y/x |  0.30                   |
| System solver     |  LAPACK dense           |
|                   |  Single precision       |
| FEAST uses MKL?   |  Yes                    |
| Fact. stored?     |  Yes                    |
| Initial Guess     |  Random                 |
| Size system       |      4                  |
| Size subspace     |      3                  |
-----------------------------------------------
 
.-------------------.
| FEAST runs        |
----------------------------------------------------------------------------------------------
#It |  #Eig  |          Trace            |     Error-Trace          |     Max-Residual
----------------------------------------------------------------------------------------------
  0      2      7.9999999999999964E+00      1.0000000000000000E+00      2.2017475048923482E-08
  1      2      7.9999999999999982E+00      3.5527136788005011E-16      1.0614112804180586E-15
 
==>FEAST has successfully converged with Residual tolerance <1E-12
   # FEAST outside it.        1
   # Eigenvalue found         2 from   3.9999999999999987E+00 to   4.0000000000000000E+00
----------------------------------------------------------------------------------------------
 
.-------------------.
| FEAST-RCI timing  |
--------------------.------------------.
| Fact. cases(10,20)|      0.0009      |
| Solve cases(11,12)|      0.0197      |
| A*x   cases(30,31)|      0.0000      |
| B*x   cases(40,41)|      0.0000      |
| Misc. time        |      0.0004      |
| Total time (s)    |      0.0210      |
--------------------------------------- 
 
***********************************************
*********** FEAST- END*************************
***********************************************
 
 Solutions (Eigenvalues/Eigenvectors/Residuals)
 E=   4.00000000000000      X= -0.409757405384329       4.544205370554897E-003
  0.814970605398106      -0.409757405384329      Res=  1.061411280418059E-015
 
 E=   4.00000000000000      X= -0.286529001556041       0.866013481533371     
 -0.292955478421287      -0.286529001556041      Res=  6.481268641478987E-016
 %\end{verbatim}
 \end{Verbatim}
\end{footnotesize}
\end{shaded}
\label{figure:hello_worldout}

\subsubsection*{C}

Similarly to the F90 example, the corresponding C source code for the helloworld example ({\tt helloworld.c}) is provided below:

\begin{small}
  \begin{C}
#include <stdio.h>
#include <stdlib.h>
#include "feast.h"
#include "feast_dense.h"
int main() {
  /*  4x4 eigenvalue system  */
  int   N=4;
  char  UPLO='F'; // 'L' and 'U' also fine
  double A[16]={2.0,-1.0,-1.0,0.0,-1.0,3.0,-1.0,-1.0,-1.0,-1.0,3.0,-1.0,0.0,-1.0,-1.0,2.0};
  /* input parameters for FEAST */
  int fpm[64];
  int    M0=3; //search subspace dimension
  double Emin=3.0, Emax=5.0; // search interval
  /* output variables for FEAST */
  double *E, *res, *X;
  double  epsout;
  int loop,info,M,i;

  /* Allocate memory for eigenvalues.eigenvectors/residual */
  E=calloc(M0,sizeof(double));  //eigenvalues
  res=calloc(M0,sizeof(double));//eigenvectors
  X=calloc(N*M0,sizeof(double));//residual

  /* !!!!!!!!!! FEAST !!!!!!!!!*/
  feastinit(fpm);
  fpm[0]=1;  /*change from default value */
  dfeast_syev(&UPLO,&N,A,&LDA,fpm,&epsout,&loop,&Emin,&Emax,&M0,E,X,&M,res,&info);

  /*!!!!!!!!!! REPORT !!!!!!!!!*/
  if (info==0) {
    printf("Solutions (Eigenvalues/Eigenvectors/Residuals)\n");
    for (i=0;i<=M-1;i=i+1){
      printf("E=%.15e X=%.15e %.15e %.15e %.15e Res=%.15e\n",
              *(E+i),*(X+i*N),*(X+1+i*N),*(X+2+i*N),*(X+3+i*N),*(res+i));
    }
  }
  return 0;
}
\end{C}
\end{small}

\noindent To create the executable, compile and link the source program with the
{\tt feast} library, one can use (for example):

\begin{itemize}  
 \item  \fbox{\parbox{\dimexpr\linewidth-2\fboxsep-2\fboxrule\relax}{\tt icc -qopenmp  -I\$FEASTROOT/include  -o helloworld helloworld.c -L\$FEASTROOT/lib/x64 -lfeast -mkl  -lirc -lifcore -lifcoremt}} \\
   if FEAST was compiled with {\tt ifort} and MKL flag was set to 'yes'.
\item  \fbox{\parbox{\dimexpr\linewidth-2\fboxsep-2\fboxrule\relax}{\tt gcc -fopenmp  -I\$FEASTROOT/include  -o helloworld helloworld.c -L\$FEASTROOT/lib/x64 -lfeast -Wl,--start-group -lmkl\_gf\_lp64 -lmkl\_gnu\_thread -lmkl\_core -Wl,--end-group -lgomp -lpthread -lm -ldl -lgfortran}} \\
   if FEAST was compiled with {\tt gfortran} and MKL flag was set to 'yes'.
   
\end{itemize}

\subsubsection*{MPI-F90}

FEAST can be straightforwardly parallelized using MPI at level L2 (where the FEAST inner linear systems are automatically
distributed among MPI processes).
As a reminder level L1 corresponds to the parallelization of the search interval,
and level L3 corresponds to the parallelization of each linear system using row-data-distribution and MPI solver.
Examples using L1-L2-L3 (MPI-MPI-MPI) are discussed in the PFEAST Section \ref{sec-pfeast}).

You can create the file {\tt phelloworld.f90} by cc-paste the content of {\tt helloworld.f90} and by just adding a few lines at the beginning and at the end of the program, i.e.

\begin{small}
  \begin{F90}
  !!!! add at the very beginning  
  include 'mpif.h'
    
  !!!! add after variable declarations
  integer :: code
  call MPI_INIT(code)


  !!!! add at the end
  call MPI_FINALIZE(code)
\end{F90}
\end{small}

\noindent Your program must be compiled using the same MPI implementation used to compile the FEAST-MPI library.
Once compiled, your source program must now be linked with the {\tt pfeast} library. You can use (for example):
\begin{itemize}
\item \fbox{\tt mpiifort -o phelloworld phelloworld.f90 -L\$FEASTROOT/lib/<{\it arch}> -lpfeast -mkl} \\
  if FEAST was compiled with {\tt ifort}, MKL flag was set to 'yes', and MPI was chosen to be 'impi' (intel mpi).
\item \fbox{\parbox{\dimexpr\linewidth-2\fboxsep-2\fboxrule\relax}{\tt
    mpif90.mpich -fc=gfortran phelloworld.f90 -o phelloworld -L\$FEASTROOT>/lib/<{\it arch}> -lpfeast -Wl,--start-group -lmkl\_gf\_lp64 -lmkl\_gnu\_thread -lmkl\_core -Wl,--end-group -lgomp -lpthread -lm -ldl -lifcore}}\\
   if FEAST was compiled with {\tt gfortran}, MKL flag was set to 'yes', and MPI was chosen to be 'mpich'.
\end{itemize}

\noindent A run of the resulting executable looks like:
         \begin{center}\fbox{\tt mpirun -ppn 1 -n <np> ./phelloworld} \end{center}
where ${\tt <np>}$ represents the number of MPI processes (here we also choose 1 MPI process per compute node with the option {\tt -ppn 1}) \\

\noindent{\bf Remarks:}\\
\noindent 1-Scalability performances will be optimal here when the number of MPI processes {\tt <np>} reaches the number of
contour points (provided by the default value {\tt fpm(2)=8} in this example). \\
\noindent 2-Since FEAST is also threaded, make sure that your number of selected threads {\tt <omp>} times the number of mpi processes {\tt <np>} for a given compute node (i.e. {\tt <omp>*<np>})  does not exceed the number of your physical cores.

\subsubsection*{MPI-C}

Similarly to the MPI-F90 example,
You can create the file {\tt phelloworld.c} by cc-paste the content of {\tt helloworld.c} and by just adding a few lines at the beginning and at the end of the program, i.e.

\begin{small}
  \begin{C}
  !!!! add at the very beginning  
  #include <mpi.h> 

  !!!! change the argument list of the main function  
  int main(int argc, char **argv)
    
  !!!! add after variable declarations
  MPI_Init(&argc,&argv);

  !!!! add at the end
  MPI_Finalize();
\end{C}
\end{small}

\noindent Your program must be compiled using the same MPI implementation used to compile the FEAST-MPI library.
Once compiled, your source program must now be linked with the {\tt pfeast} library. You can use (for example):
\begin{itemize}
\item  \fbox{\parbox{\dimexpr\linewidth-2\fboxsep-2\fboxrule\relax}{\tt mpiicc -qopenmp  -I\$FEASTROOT/include  -o phelloworld phelloworld.c -L\$FEASTROOT/lib/x64 -lpfeast -mkl  -lirc -lifcore -lifcoremt}} \\
  if FEAST was compiled with {\tt ifort},
  MKL flag was set to 'yes',  and MPI was chosen to be 'impi' (intel mpi).

\item  \fbox{\parbox{\dimexpr\linewidth-2\fboxsep-2\fboxrule\relax}{\tt mpicc -cc=gcc -fopenmp  -I\$FEASTROOT/include  -o phelloworld phelloworld.c -L\$FEASTROOT/lib/x64 -lpfeast -Wl,--start-group -lmkl\_gf\_lp64 -lmkl\_gnu\_thread -lmkl\_core -Wl,--end-group -lgomp -lpthread -lm -ldl -lgfortran}} \\
  if FEAST was compiled with {\tt gfortran},
  MKL flag was set to 'yes',  and MPI was chosen to be 'mpich'.   
\end{itemize}

\newpage






\section{FEAST Interfaces}

\subsection{At a Glance}

There are the two different types of interfaces available in the FEAST library:\\

\noindent \fbox{\bf Driver interfaces}
\begin{itemize}
\itemsep 1pt
\parskip 1pt
\item Optimal drivers acting on  commonly used matrix data storage (dense, banded, sparse-CSR, row-distributed CSR).
\item Use predefined linear system solvers: LAPACK (for dense), SPIKE (for banded), MKL-PARDISO (for sparse-CSR and FEAST),
BiCGStab (for sparse-CSR and IFEAST), MKL-CLUSTER-PARDISO (for row-distributed CSR and PFEAST), PBiCGStab (for row-distributed CSR and
PIFEAST).
\end{itemize}

\noindent \fbox{\bf Reverse communication interfaces (RCI)}
\begin{itemize}
\itemsep 1pt
\parskip 1pt
\item Constitute the kernel of FEAST, independent of the matrix data formats, so users
  can easily customize FEAST with their own explicit or implicit data format (or row-distributed data format). 
\item  Mat-vec routines and
  direct/iterative linear system solvers must also be provided by the users.
  \end{itemize}


\noindent Here is the complete list of all FEAST v4.0 interfaces (186 in total).

\begin{table}[h!]
\begin{small}
  \begin{center}
  \begin{tabular}{|l||c|c|c|} 
    \hline
          Properties  & RCI interfaces & Dense/{\color{blue}{Banded}} interfaces & Sparse interfaces\\ \hline \hline
  \col {\bf Linear  $AX=BX\Lambda$} &  \multicolumn{3}{c|}{\col} \\ \hline
 Real Sym. $A=A^T$,$B$ spd & \tt dfeast\_srci\{x\} & \tt dfeast\_\{sy,{\color{blue} sb}\}\{ev,gv\}\{x\}  &\tt \{p\}d\{i\}feast\_scsr\{ev,gv\}\{x\}\\  
       Complex Herm.   $A=A^H$,$B$ hpd
         & \tt zfeast\_hrci\{x\} &  \tt zfeast\_\{he,{\color{blue} hb}\}\{ev,gv\}\{x\} & \tt \{p\}z\{i\}feast\_hcsr\{ev,gv\}\{x\}  \\ 
       Complex  Sym.  $A=A^T$,$B=B^T$
       & \tt zfeast\_srci\{x\} & \tt zfeast\_\{sy,{\color{blue} sb}\}\{ev,gv\}\{x\} &  \tt \{p\}z\{i\}feast\_scsr\{ev,gv\}\{x\}  \\
       Real  General & \tt zfeast\_grci\{x\} &  \tt dfeast\_\{ge,{\color{blue} gb}\}\{ev,gv\}\{x\} &  \tt \{p\}d\{i\}feast\_gcsr\{ev,gv\}\{x\} \\ 
       Complex  General  & \tt zfeast\_grci\{x\} &   \tt zfeast\_\{ge,{\color{blue} gb}\}\{ev,gv\}\{x\} &  \tt \{p\}z\{i\}feast\_gcsr\{ev,gv\}\{x\} \\ \hline
        \col {\bf Polynomial} $\sum_{i}{A_i}X\Lambda^i=0$ &  \multicolumn{3}{c|}{\col}  \\ \hline
        Real Sym. $A_i=A_i^T$ & \tt zfeast\_srcipev\{x\} &   \tt dfeast\_sypev\{x\} &  \tt \{p\}d\{i\}feast\_scsrpev\{x\} \\
        Complex Herm. $A_i=A_i^H$  & \tt zfeast\_grcipev\{x\} &   \tt zfeast\_hepev\{x\} &  \tt \{p\}z\{i\}feast\_hcsrpev\{x\} \\
        Complex Sym. $A_i=A_i^T$  & \tt zfeast\_srcipev\{x\} &   \tt zfeast\_sypev\{x\} &  \tt \{p\}z\{i\}feast\_scsrpev\{x\} \\
        Real General  & \tt zfeast\_grcipev\{x\} &   \tt dfeast\_gepev\{x\} &  \tt \{p\}d\{i\}feast\_gcsrpev\{x\} \\
 Complex General. & \tt zfeast\_grcipev\{x\} &   \tt zfeast\_gepev\{x\} &  \tt \{p\}z\{i\}feast\_gcsrpev\{x\} \\ \hline 
\end{tabular}
\end{center}
\end{small}
\end{table}

\noindent where 

\begin{itemize}
\itemsep 1pt
\parskip 1pt
\item {\tt dfeast} and  {\tt zfeast} stand for real double precision and complex double precision, respectively. 
\item {\tt \{ev,gv\}} stands for either standard (i.e. {\tt B=I}) or generalized eigenvalue problems.
\item {\tt \{x\}} is optional - stands for the expert FEAST version which enables customized quadrature nodes/weights.
\item {\tt \{i\}} is optional - stands for the IFEAST version of the sparse interfaces using inexact iterative solver.
\item {\tt \{p\}} is optional - stands for the PFEAST version of the sparse interfaces using distributed MPI solvers (direct or iterative).
\end{itemize}

\noindent In addition, all the input parameters for the FEAST algorithm are contained into an integer array
of size 64 named here {\tt fpm}. Prior calling the FEAST interfaces, this array needs to be initialized. There exists two FEAST initialization routines:

\begin{framed}
\begin{tt}
\noindent {\color{mygreen}{!!!! initialization for FEAST}}\\[3pt]
\noindent feastinit(fpm)\\[3pt]
\noindent {\color{mygreen}{!!!! initialization for PFEAST (needed if the sparse interfaces use a MPI solver)}}\\[3pt]
\noindent pfeastinit(fpm,L1\_comm\_world,nL3) 
\end{tt}
\end{framed}

\noindent All input FEAST parameters are then set to their default values. The detailed list of the {\tt fpm} parameters is given in
Table  \ref{tab_fpm}. Users can modify their values accordingly before calling the FEAST interfaces.

\begin{table}[htbp]
\begin{center}
\begin{small}
\begin{tabular}{|c|p{8.5cm}|p{4.6cm}|}
\hline
{\tt fpm(i)} F90 &  ~~~~~~~~~~~~~~~~~~~~~~~~~~~~~~~~Description & ~~~~~~Default values \\ 
{\tt fpm[i-1]} C &  &  \\ \hline  \hline
\multicolumn{3}{|c|}{\col Runtime and algorithm options} \\ \hline
{\tt i=1}  & 
Print runtime comments \newline
0: Off || 1: On screen \newline n<0: Write/Append comments in the file {\tt feast<|n|>.log} & 0 \\  \hline
{\tt i=2}  & \#contour points for Hermitian FEAST (half-contour) \newline
if {\tt fpm(16)=0,2}, values permitted (1 to 20, 24, 32, 40, 48, 56) \newline
if {\tt fpm(16)=1}, all positive values permitted& 8 using FEAST \newline 4 using IFEAST \newline 3 using Stochastic {\tt fpm(14)=2}\\ \hline
{\tt i=3}  &  Stopping convergence criteria in double precision (0 to 16)\newline $\epsilon=10^{\tt -fpm(3)}$  & 12 \\ \hline
{\tt i=4}  &  Maximum number of FEAST refinement loop allowed ($\geq 0$)  & 20 using FEAST \newline 50 using IFEAST\\ \hline
{\tt i=5}  &  Provide initial guess subspace (0: No; 1: Yes) & 0 \\ \hline
{\tt i=6}  &  Convergence criteria (for solutions  in the search contour)\newline
0: Using relative error on the trace {\tt epsout} i.e. {\tt epsout$< \epsilon$} \newline
1: Using relative residual {\tt res} i.e. $\max_i {\tt res(i)}< \epsilon$ & 1  \\  \hline
{\tt i=8}  & \#contour points for non-Herm./poly. FEAST (full-contour) \newline
 if {\tt fpm(16)=0}, values permitted (2 to 40, 48, 64, 80, 96, 112) \newline
if {\tt fpm(16)=1}, all values permitted ($>$2) & 16 using FEAST \newline  8 using IFEAST \newline 6 using Stochastic {\tt fpm(14)=2} \\  \hline
{\tt i=9}  &  L2 communicator for PFEAST & set by call to {\tt pfeastinit} \\ \hline
{\tt i=10}  &  Store linear system factorizations  (0: No; 1: Yes).  & 1 using all Driver interfaces \newline 0 using RCI interfaces \\ \hline
{\tt i=14} &  0: FEAST normal execution \newline 1: Return subspace Q after 1 contour \newline
 2: Stochastic estimate of \#eigenvalues inside search contour & 0  \\  \hline
{\tt i=15} &   \#Contours for non-Hermitian or polynomial FEAST. \newline
0: two-sided contour (compute right/left eigenvectors) \newline 1: one-sided contour (compute only right eigenvectors)
\newline 2: one sided contour (left=right* eigenvectors) & 0 using non-sym. drivers \newline  1 using Stochastic {\tt fpm(14)=2} \newline
2 using sym. drivers\\ \hline 
{\tt i=16} &   Integration type (0: Gauss 1: Trapezoidal; 2: Zolotarev) \newline Remark: option 2 only for Hermitian& 0 for Hermitian FEAST \newline 1 for non-Herm./poly. FEAST \newline 1 for IFEAST\\ \hline
{\tt i=18} &   Ellipse contour ratio  'vertical axis'/'horizontal axis' ($\geq 0$)\newline  {\tt fpm(18)}/100 = ratio & 30 for Hermitian FEAST\newline 100 for non-Herm./poly. FEAST \newline 100 for IFEAST \\ \hline
{\tt i=19} &   Ellipse rotation angle in degree from vertical axis [-180:180]
\newline Remark: only for non-Hermitian & 0 \\\hline
{\tt i=49}  &  L3 communicator for PFEAST & set by call to {\tt pfeastinit} \\ \hline
\multicolumn{3}{|c|}{\col Driver interface options} \\ \hline 
{\tt i=40} &  Search interval option for sparse Hermitian drivers\newline
     0: search interval provided by user \newline
    -1: search M0/2 lowest eigenvalues- return search interval \newline
~1: search M0/2 largest eigenvalues- return search interval & 0 \\ \hline
{\tt i=41}  &  Matrix scaling for sparse drivers (0: No; 1: Yes).  & 1 \\ \hline
{\tt i=42}  &  Mixed Precision for all drivers\newline 0: use double precision linear system solvers \newline 1: use single precision linear system solvers & 1 \\ \hline
{\tt i=43}  &  Automatic switch from FEAST to IFEAST drivers \newline (0:{\tt feast},1:{\tt ifeast}) & 0 \\ \hline
{\tt i=45}  &  Accuracy of BiCGStab  in IFEAST  $\mu=10^{\tt -fpm(45)}$ & 1 \\ \hline
{\tt i=46}  &  Maximum \#iterations for BiCGStab in IFEAST & 40 \\ \hline
{\tt i=60}  &  {\bf Output:}  returns the total number of BicGstab iterations  & N/A \\ \hline
{\tt All Others}  & Reserved values and/or Undocumented options & N/A \\ \hline
\end{tabular}
\caption{\label{tab_fpm}
List  FEAST parameters with default input values. \newline
{\bf Remark:} Using the {\tt C} language, the components of the {\tt fpm} array starts at 0 and stops at 63.
Therefore, the components {\tt fpm[j]} in C ({\tt j=0-63}) must correspond to the components {\tt fpm(i)}
in Fortran ({\tt i=1-64}) specified above (i.e. {\tt fpm[i-1]}={\tt fpm(i)}).}
\end{small}
\end{center}
\end{table}

\newpage

\noindent Errors and warnings encountered during a run of the FEAST package
are stored in an integer variable, {\tt info}. If the value of the output {\tt info} parameter
is different than ``0'', either an error or warning  was encountered.
The possible return values for the {\tt info} parameter along with the error code descriptions, 
are given in Table~\ref{tab_info}.

\begin{table}[htbp]
\begin{small}
\begin{center}
\begin{tabular}{|l l p{4.6in}|}
\hline
{\tt info} & Classification & Description \\\hline\hline
$202$ & Error & Problem with size of the system {\tt N} \\ 
$201$ & Error & Problem with size of subspace {\tt M0} \\ 
$200$ & Error & Problem with Emin, Emax or Emid, r \\ 
$(100+i)$ & Error & Problem with $i^{th}$ value of the input FEAST parameter (i.e {\tt fpm(i)}) \\
$7$    & Warning &  The search for extreme eigenvalues has failed, search contour must be set by user\\
$6$    & Warning &  FEAST converges but subspace is not bi-orthonormal \\
$5$   & Warning & Only stochastic estimation of \#eigenvalues returned {\tt fpm(14)=2} \\
$4$    & Warning & Only the subspace has been returned using {\tt fpm(14)=1} \\
$3$    & Warning & Size of the subspace {\tt M0} is too small ({\tt M0<=M}) \\
$2$    & Warning & No Convergence (\#iteration loops$>${\tt fpm(4)})\\
$1$    & Warning & No Eigenvalue found in the search interval    \\ \hline
\col $0$    &  \col Successful exit& \col\\ \hline
$-1$ & Error & Internal error conversion single/double \\
$-2$ &  Error    & Internal error of the inner system solver in FEAST Driver interfaces \\
$-3$ &  Error    & Internal error of the reduced eigenvalue solver \\
     &           & {\it Possible cause for Hermitian problem: matrix {\bf B} may not be positive definite} \\
$-(100+i)$    & Error & Problem with the $i^{th}$ argument of the FEAST interface\\
\hline
\end{tabular}
\caption{Return code descriptions for the parameter {\tt info}.}
\label{tab_info}
\end{center}
\end{small}
\end{table}


\subsubsection*{\underline{Quick Tutorial using FEAST Drivers}}
\begin{enumerate}
\itemsep 1pt
\parskip 1pt
\item Identify your FEAST driver. Look at the corresponding section
  of this documentation to set up the argument lists.
\item Specify a search contour enclosing the wanted eigenvalues (normal FEAST mode). Alternatively, you can also use the Hermitian
sparse drivers to search for the {\tt M0/2} lowest ({\tt fpm(40)=-1}) or largest  ({\tt fpm(40)=1}) eigenpairs (here {\tt M0} must be set to
two times the number of wanted eigenvalues).
\item In normal FEAST mode, specify the search subspace size {\tt M0} as an overestimation of your estimated \#eigenvalues {\tt M} within the contour (typically {\tt M0}$\geq$ 1.5{\tt M}). If needed, user can take advantage
of fast stochastic estimates for {\tt M} within a particular contour using {\tt fpm(14)=2}. 
\item Change the {\tt fpm} default options if needed and run the code.
\end{enumerate}

\subsubsection*{\underline{Tips}}

\begin{itemize}
\itemsep 1pt
\parskip 1pt

\item  The FEAST convergence rate depends on the choice of the search subspace size {\tt M0},
the number of contour points, and the nature of the quadrature.
To  improve the convergence rate, you have the possibility to:
\begin{itemize}
\itemsep 1pt
\parskip 1pt
\item keep on increasing the number of quadrature nodes {\tt fpm(2)}
 ({\tt fpm(8)} for non-Hermitian/Polynomial).
\item keep on increasing {\tt M0} for the Gauss-Legendre or Trapezoidal quadrature.
\item switch to Zolotarev quadrature for the Hermitian problem with {\tt fpm(16)=2}  (Zolotarev is ideally suited to deal with continuum spectra without the need to increase  the subspace size {\tt M0$\sim$M}).
\end{itemize}

\item Although FEAST could be used to seek $1000$'s of
eigenpairs within a single search contour, the size of the search subspace {\tt M0} is
supposed to be much smaller than the size of the eigenvalue problem  {\tt N}. As a
result, the arithmetic complexity would mainly depend on the inner system solve
(i.e. $\rm O(NM_0)$ for narrow banded or sparse system solvers).
If you are looking for a very large number of eigenvalues, it is recommended to consider
multiple search intervals to be solved in parallel using PFEAST.

\item FEAST v4.0 is using an inverse residual iteration algorithm which enables the linear
systems to be solved with very low accuracy with no impact on the FEAST double precision
convergence rate (!). Consequently, all FEAST linear systems are solved in single precision
by default ({\tt fpm(42)=1}). Using the RCI interfaces, users can then plug in their own low accuracy (single precision or less)
direct or iterative solver. Additionally, all the  linear system factorizations
can be kept in memory by  (using {\tt fpm(10)=1}) which improves performance but use more memory.

\item Using the FEAST-SMP library, parallelism at the third level L3 (linear system solves) can only be achieved using the threading capabilities of
the linear system solver and via the shell variable {\tt MKL\_NUM\_THREADS} if Intel-MKL is used or
the  shell variable {\tt OMP\_NUM\_THREADS} if
SPIKE is used for the banded interfaces.



\item Using the FEAST-MPI library, you can trivially parallelize the second level L2 (contour points) and keep on using
the same FEAST/IFEAST driver interfaces with shared memory solver at L3. Scalability performances will be optimal when the number of MPI processes reaches the number of contour points (either {\tt fpm(2)} for FEAST Hermitian or {\tt fpm(8)} for FEAST non-Hermitian and Polynomial).

\item The FEAST-MPI library also offers the possibility to use a MPI solver at level L3. This scheme is called PFEAST and it is detailed in Section \ref{sec-pfeast}.

\end{itemize}


\newpage

\subsection{FEAST Hermitian}\label{sec-Hermitian}


\begin{tt}
\noindent {\color{mygreen}{!!!! Standard AX=EX - Real-Symmetric and Complex Hermitian}}\\[3pt]
\noindent dfeast\_s{\large \bf F}ev{\color{blue}{\{x\}}}({\bf \{List-A\}},fpm,epsout,loop,Emin,Emax,M0,E,X,M,res,info,{\color{blue}{\{Zne,Wne\}}})\\[3pt]
\noindent zfeast\_h{\large \bf F}ev{\color{blue}{\{x\}}}({\bf \{List-A\}},fpm,epsout,loop,Emin,Emax,M0,E,X,M,res,info,{\color{blue}{\{Zne,Wne\}}}) \\[3pt]
\noindent {\color{mygreen}{!!!! Generalized AX=EBX - Real-Symmetric and Complex Hermitian- (B is hpd)}}\\[3pt]
\noindent dfeast\_s{\large \bf F}gv{\color{blue}{\{x\}}}({\bf \{List-A\}},{\bf \{List-B\}},fpm,epsout,loop,Emin,Emax,M0,E,X,M,res,info,{\color{blue}{\{Zne,Wne\}}})\\[3pt]
\noindent zfeast\_h{\large \bf F}gv{\color{blue}{\{x\}}}({\bf \{List-A\}},{\bf \{List-B\}},fpm,epsout,loop,Emin,Emax,M0,E,X,M,res,info,{\color{blue}{\{Zne,Wne\}}})\\[3pt]
\noindent {\color{mygreen}{!!!! RCI (format independent) - Real-Symmetric and Complex Hermitian}}\\[3pt]
\noindent dfeast\_srci{\color{blue}{\{x\}}}({\color{red}{ijob,N,Ze,work1,work2,Aq,Bq}},fpm,epsout,loop,Emin,Emax,M0,E,X,M,res,info,{\color{blue}{\{Zne,Wne\}}})\\[3pt]
\noindent zfeast\_hrci{\color{blue}{\{x\}}}({\color{red}{ijob,N,Ze,work1,work2,Aq,Bq}},fpm,epsout,loop,Emin,Emax,M0,E,X,M,res,info,{\color{blue}{\{Zne,Wne\}}})\\[3pt]
\end{tt}

\noindent We note the following:
\begin{itemize}
\itemsep 1pt
\parskip 1pt
\item The Table below details the series of arguments in each {\tt \bf \{List-A\}}, and {\tt \bf \{List-B\}} that are specific to the type of matrix format represented above by
{\large \tt \bf F} (as a placeholder). 
\begin{center}
\begin{small}
\begin{tabular}{|c||c|c|c|}
\hline
\tt  & {\tt \bf F} & {\tt \bf List-A} & {\tt \bf List-B}  \\
\hline
\hline
\multicolumn{1}{|c||}{\col \tt Dense} & \tt \{y,e\} &\tt  \{ UPLO, N, A, LDA \} &\tt  \{ B, LDB \} \\ \hline
\multicolumn{1}{|c||}{\col \tt Banded} & \tt b &\tt  \{ UPLO, N, ka, A, LDA \} &\tt  \{ kb, B, LDB \}\\ \hline
\multicolumn{1}{|c||}{\col \tt  Sparse} &\tt csr &\tt  \{ UPLO, N, A, IA, JA \} &\tt  \{ B, IB, JB \}  \\ \hline
\end{tabular}
\end{small}
\end{center}

\item  Table \ref{tab-spec1} details the specific matrix-format arguments in {\tt \bf \{List-A\}} and {\tt \bf \{List-B\}}
\item Table \ref{tab-common1} details the common arguments in all the Hermitian FEAST interfaces above,
\item Table \ref{tab-rci1} details the arguments for the Hermitian RCI interfaces (in red above). 
\end{itemize}

\begin{table}[h!]
\begin{center}
\begin{small}
\begin{tabular}{|l|l|c|l|}
\hline
 & \multicolumn{1}{|c|}{Type} & I/O &\multicolumn{1}{|c|}{Description} \\ \hline \hline
{\tt fpm}  & 
integer(64)  & in/out & FEAST input parameters (see Table \ref{tab_fpm})\\ \hline
{\tt epsout}  & double real  & out & Trace relative error {\footnotesize $|{\tt trace}_k-{\tt trace}_{k-1}|/\max({\tt |Emin|,|Emax|})$}\\ \hline
{\tt loop} &  integer           &  out            &    \# of FEAST subspace iterations\\ \hline
{\tt Emin,Emax}  &   double real   &  in/out  & Lower and Upper bounds of search interval\\
 &    &    & Remark: Output values if {\tt fpm(40)=+-1} for sparse drivers\\ \hline
{\tt M0}   &    integer         &    in/out          &  Search subspace dimension               \\ 
&                    &                          & On entry: initial guess {\tt M0$\geq$M} \\
&                    &                          & On exit:  new suitable {\tt M0} if guess too large    \\
&                    &                          & Remark: {\tt M0=2*Wanted} if {\tt fpm(40)=+-1} for sparse drivers       \\ \hline
{\tt E}    &  double real({\tt M0})      &     in/out         & Eigenvalues    \\
&                              &                   &  On entry: initial guess if {\tt fpm(5)=1} (previous FEAST run) \\
&                              &                   &  On exit: Eigenvalues  solutions {\tt E(1:M)}   \\
           &                   &                 &  Remark: the {\tt E(M+1:M0)} values are outside {\tt [Emin,Emax]}\\\hline 
{\tt X}    &   double real({\tt N,M0}) using {\tt dfeast}           &     in/out      & Eigenvectors ({\tt N}: size of the system)                \\
           &   double complex({\tt N,M0}) using {\tt zfeast}         &                & On entry: initial guess if {\tt fpm(5)=1}  (previous FEAST run) \\
           &             &                   & On exit: Eigenvectors solutions  {\tt X(1:N,1:M)}\\
           &  &  & Remark: if {\tt fpm(14)=1}, first $\bf Q$ subspace on exit  \\ \hline
{\tt M} &     integer        &     out         & \#Eigenvalues found in   {\tt [Emin,Emax]}   \\ 
 &             &             & \#Estimated eigenvalues if {\tt fpm(14)=2} \\  \hline
{\tt res} &  double real({\tt M0})          &   out     & Relative residual   $\bf ||Ax_i - \lambda_i Bx_i ||_2 /||\alpha Bx_i ||_2$\\
&   &   & with ${\tt \alpha}=\max({\tt |Emin|,|Emax|})$   \\ \hline
{\tt info} &  integer        &  out           &   Error handling  (see Table \ref{tab_info} for all INFO codes) \\ \hline
{\tt Zne,Wne}  &   double complex({\tt fpm(2)})    &  in  & Custom integration nodes and weights- Expert mode\\ \hline
\end{tabular}
\caption{\label{tab-common1} List of common arguments for the FEAST Hermitian Driver interfaces. 
}
\end{small}
\end{center}
\end{table}

\begin{table}[htbp]
\begin{center}
\begin{small}
\begin{tabular}{|l|p{6cm}|c|l|}
\hline
 & \multicolumn{1}{|c|}{Type} & I/O &\multicolumn{1}{|c|}{Description}  \\ 
\hline  \hline
   \multicolumn{4}{|l|}{\col Common}    \\ \hline
{\tt UPLO}  & 
character(len=1) & in & Matrix Storage {\tt ('F','L','U')} \\ 
& & & 'F': Full; 'L': Lower; 'U': Upper\\ \hline
{\tt N}  & integer  & in & Size of the system \\ \hline
\multicolumn{4}{|l|}{\col Dense}   \\ \hline
{\tt A}  &  double real({\tt LDA,N}) \hfill using {\tt dfeast}    &     in       
& Eigenvalue system (Stiffness) matrix         \\
  &  double complex({\tt LDA,N})  \hfill using {\tt zfeast}    &           
&          \\ \hline
{\tt LDA}  &    integer          &     in       
& Leading dimension of {\tt A} {\tt LDA>=N}  \\ \hline
{\tt B}  &  double real({\tt LDB,N})  \hfill using {\tt dfeast}    &     in       
& Eigenvalue system (Mass) matrix         \\
  &  double complex({\tt LDA,N})  \hfill using {\tt zfeast}    &           
&          \\ \hline
{\tt LDB}  &    integer          &     in       
& Leading dimension of {\tt B} {\tt LDB>=N}  \\ \hline
\multicolumn{4}{|l|}{\col Banded}   \\ \hline
{\tt ka}  &    integer          &     in       
& The number of sub or super-diagonals  within the band of {\tt A}.  \\ \hline
{\tt A}  &  double real({\tt LDA,N})  \hfill using {\tt dfeast}    &     in       
& Eigenvalue system (Stiffness) matrix         \\
  &  double complex({\tt LDA,N})  \hfill using {\tt zfeast}    &           
&          \\ \hline
{\tt LDA}  &    integer          &     in       
& Leading dimension of {\tt A} {\tt LDA>=2*ka+1} if  {\tt UPLO='F'}\\
& & & {\color{white}{Leading dimension of {\tt A}}} {\tt LDA>=ka+1} if  {\tt UPLO='L' or 'U'}   \\ \hline
{\tt kb}  &    integer          &     in       
& The number of sub or super-diagonals  within the band of {\tt B}.  \\ \hline
{\tt B}  &  double real({\tt LDB,N})  \hfill using {\tt dfeast}    &     in       
& Eigenvalue system (Mass) matrix         \\
  &  double complex({\tt LDB,N})  \hfill using {\tt zfeast}    &           
&          \\ \hline
{\tt LDB}  &    integer          &     in       
& Leading dimension of {\tt B} {\tt LDB>=2*kb+1} if  {\tt UPLO='F'}\\
& & & {\color{white}{Leading dimension of {\tt B}}} {\tt LDB>=kb+1} if  {\tt UPLO='L' or 'U'}   \\ \hline
\multicolumn{4}{|l|}{\col Sparse-csr}   \\ \hline
{\tt A}  &  double real({\tt IA(N+1)-1})  \hfill using {\tt dfeast}    &     in       
& Eigenvalue system (Stiffness) matrix - CSR values         \\
  &  double complex({\tt IA(N+1)-1})  \hfill using {\tt zfeast}    &           
&          \\ \hline
{\tt IA}  &   integer(N+1)          &     in       
  & Sparse CSR Row array of {\tt A}.          \\ \hline
{\tt JA}  &   integer(IA(N+1)-1)          &     in       
  & Sparse CSR Column array of {\tt A}.          \\ \hline
{\tt B}  &  double real({\tt IB(N+1)-1})  \hfill using {\tt dfeast}    &     in       
& Eigenvalue system (Mass) matrix - CSR values         \\
  &  double complex({\tt IB(N+1)-1})  \hfill using {\tt zfeast}    &           
&          \\ \hline
{\tt IB}  &   integer(N+1)          &     in       
  & Sparse CSR Row array of {\tt B}.          \\ \hline
{\tt JB}  &   integer(IB(N+1)-1)          &     in       
  & Sparse CSR Column array of {\tt B}.          \\ \hline
\end{tabular}
\end{small}
\caption{\label{tab-spec1}   List of arguments that are matrix-format specific for the FEAST  Driver interfaces. Applicable to Hermitian and Non-Hermitian Drivers.}
\end{center}
\end{table}

\begin{table}[htbp]
\begin{center}
\begin{small}
\begin{tabular}{|l|p{5.5cm}|c|l|}
\hline
 & \multicolumn{1}{|c|}{Type} & I/O &\multicolumn{1}{|c|}{Description}  \\ 
\hline  \hline
{\tt ijob}  & 
integer  & in/out &  On entry: ijob=-1 (initialization)\\ 
& & & On exit:  ID of the FEAST\_RCI operation \\ \hline
{\tt N}  & integer  & in & Size of the system \\ \hline
{\tt Ze} & double complex   &  out    &  Coordinate along the complex contour  \\ \hline
{\tt work1} &   double real({\tt N,M0}) \hfill using {\tt dfeast}    &     in/out       
  & Workspace               \\
            &  double complex({\tt N,M0}) \hfill using {\tt zfeast} & &  \\ \hline 
{\tt work2} &  double complex({\tt N,M0})   &  in/out    &  Workspace  \\  \hline
{\tt Aq, Bq} &  double real({\tt M0,M0}) \hfill using {\tt dfeast}   &     in/out       
& Workspace for the reduced eigenvalue problem         \\
            &  double complex({\tt M0,M0}) \hfill using {\tt zfeast} & &  \\ \hline 
\end{tabular}
\caption{\label{tab-rci1} List of arguments for the FEAST Hermitian RCI interfaces.}
\end{small}
\end{center}
\end{table}

\newpage

\subsubsection*{\underline{Hermitian Driver Interfaces: Examples}}

Let us consider the following systems:

\begin{description}
\item[System1] a ``real symmetric'' generalized eigenvalue problem $\bf Ax=\lambda Bx$ , where  
$\bf A$ is real symmetric and $\bf B$ is symmetric positive definite.
$\bf A$ and $\bf B$ are of the size $\rm N=1671$ and
have the same sparsity pattern with number of non-zero elements $\rm NNZ=11435$.  
\item[System2] a ``complex Hermitian'' standard eigenvalue problem $\bf Ax=\lambda x$, where  
$\bf A$ is complex Hermitian. 
$\bf A$ is of size $\rm N=600$ with number of non-zero elements $\rm NNZ=2988$.
\end{description}

The {\tt \$FEASTROOT/example/FEAST} directory provides Fortran and C implementation of these systems using both dense, banded and sparse-CSR storage. Here, the complete list of routines:

\begin{table}[h!]
\begin{center}
\begin{tabular}{lll}
& \multicolumn{1}{c}{System1} &  \multicolumn{1}{c}{System2} \\ \hline\hline
\col dense & \col & \col \\
&{\tt \{F90,C\}dense\_dfeast\_sygv} &  {\tt \{F90,C\}dense\_zfeast\_heev} \\
\col banded & \col & \col \\
&{\tt \{F90,C\}dense\_dfeast\_sbgv} &  {\tt \{F90,C\}dense\_zfeast\_hbev} \\
\col sparse & \col & \col \\
&{\tt \{F90,C\}dense\_dfeast\_scsrgv}  & {\tt \{F90,C\}dense\_zfeast\_hcsrev} \\
& {\tt \{F90,C\}dense\_dfeast\_scsrgv\_lowest}$^*$ & \\ 
& ({\em $^*$using {\tt dfeast\_scsrgv} to compute few lowest eig.}) & \\ \hline
\end{tabular}
\end{center}
\end{table}

The {\tt \$FEASTROOT/example/PFEAST-L2} directory provides
the parallel implementation of all these routines using FEAST-MPI.
The routine names are preceded by the letter {\tt P}. 
The MPI parallelization operates only at the second level L2 where all the ({\tt fpm(2)}) linear systems are distributed among the MPI processes.

\subsubsection*{\underline{Hermitian RCI Interfaces}}

Using the FEAST\_RCI interfaces, the {\tt ijob} parameter must first be  initialized with the value $-1$. Once the RCI interface is called, the value of the {\tt ijob} output parameter, if different than $0$,  is used to identify the FEAST operation that needs to be completed
by the user. Users have then the possibility to customize their own matrix direct or iterative factorization and linear solve 
techniques as well as their own matrix multiplication routine.

Here is a general (F90) template example of RCI for solving real symmetric problem:


\begin{small}
\begin{F90}%[frame=trBL]
ijob=-1 ! initialization
do while (ijob/=0)
call dfeast_srci(ijob,N,Ze,work1,work2,Aq,Bq,fpm,epsout,loop,Emin,Emax,M0,E,X,M,res,info)
 select case(ijob)
 case(10) !!Factorize the complex matrix Az <=(ZeB-A) - or factorize a preconditioner of ZeB-A
          !!REMARK:  Az can be formed and factorized using single precision arithmetic
................ <<< user entry
 case(11) !!Solve the linear system with fpm(23) rhs; Az * Qz=work2(1:N,1:fpm(23)) 
          !!Result (in place) in work2 <= Qz(1:N,1:fpm(23))
          !!REMARKS:  -Solve can be performed in single precision
          !!          -Low accuracy iterative solver are ok          
................ <<< user entry
 case(30) !!Perform multiplication A * X(1:N,i:j) result in work1(1:N,i:j)
          !!        where i=fpm(24) and j=fpm(24)+fpm(25)-1
................ <<< user entry
 case(40) !!Perform multiplication B * X(1:N,i:j) result in work1(1:N,i:j)
          !!         where i=fpm(24) and j=fpm(24)+fpm(25)-1
          !!REMARK: user must set work1(1:N,i:j)=X(1:N,i:j) if B=I
................  <<< user entry
 end select
end do
 \end{F90}
\end{small}

Here is a general (F90) template example of RCI for solving complex Hermitian problem:

\begin{small}
\begin{F90}%[frame=trBL]
ijob=-1 ! initialization
do while (ijob/=0)
call zfeast_hrci(ijob,N,Ze,work1,work2,Aq,Bq,fpm,epsout,loop,Emin,Emax,M0,E,X,M,res,info)
 select case(ijob)
 case(10) !!Factorize the complex matrix Az <=(ZeB-A) - or factorize a preconditioner of ZeB-A
          !!REMARK:  Az can be formed and factorized using single precision arithmetic
................ <<< user entry
 case(11) !!Solve the linear system with fpm(23) rhs; Az * Qz=work2(1:N,1:fpm(23)) 
          !!Result (in place) in work2 <= Qz(1:N,1:fpm(23))
          !!REMARKS:  -Solve can be performed in single precision
          !!          -Low accuracy iterative solver are ok
................ <<< user entry
 case(20) !![Optional: *only if* needed by case(21)]
          !!Factorize the complex matrix Az^H
          !!REMARKS: -The matrix Az from case(10) cannot be overwritten  
          !!         -case(20) becomes obsolete if the solve in case(21) can be performed  
          !!           by reusing the factorization in case(10)
................ <<< user entry
 case(21) !!Solve the linear system with fpm(23) rhs;  Az^H * Qz=work2(1:N,1:fpm(23))
          !!Result (in place) in work2 <= Qz(1:N,1:fpm(23))          
................ <<< user entry
 case(30) !!Perform multiplication A * X(1:N,i:j) result in work1(1:N,i:j)
          !!         where i=fpm(24) and j=fpm(24)+fpm(25)-1
................ <<< user entry
 case(40) !!Perform multiplication B * X(1:N,i:j) result in work1(1:N,i:j)
          !!        where i=fpm(24) and j=fpm(24)+fpm(25)-1
          !!REMARK: user must set work1(1:N,i:j)=X(1:N,i:j) if B=I
................  <<< user entry
 end select
end do
 \end{F90}
\end{small}


\newpage

\subsection{FEAST Non-Hermitian}


\begin{tt}
\noindent {\color{mygreen}{!!!! Standard AX=EX - Complex Symmetric, Real General and Complex General}}\\[3pt]
\noindent zfeast\_s{\large \bf F}ev{\color{blue}{\{x\}}}({\bf \{List-A\}},fpm,epsout,loop,Emid,r,M0,E,X,M,res,info,{\color{blue}{\{Zne,Wne\}}})\\[3pt]
\noindent dfeast\_g{\large \bf F}ev{\color{blue}{\{x\}}}({\bf \{List-A\}},fpm,epsout,loop,Emid,r,M0,E,X,M,res,info,{\color{blue}{\{Zne,Wne\}}}) \\[3pt]
\noindent zfeast\_g{\large \bf F}ev{\color{blue}{\{x\}}}({\bf \{List-A\}},fpm,epsout,loop,Emid,r,M0,E,X,M,res,info,{\color{blue}{\{Zne,Wne\}}}) \\[3pt]
\noindent {\color{mygreen}{!!!! Generalized AX=EBX -  Complex Symmetric (B also sym.), Real General and Complex General}}\\[3pt]
\noindent zfeast\_s{\large \bf F}gv{\color{blue}{\{x\}}}({\bf \{List-A\}},{\bf \{List-B\}},fpm,epsout,loop,Emid,r,M0,E,X,M,res,info,{\color{blue}{\{Zne,Wne\}}})\\[3pt]
\noindent dfeast\_g{\large \bf F}gv{\color{blue}{\{x\}}}({\bf \{List-A\}},{\bf \{List-B\}},fpm,epsout,loop,Emid,r,M0,E,X,M,res,info,{\color{blue}{\{Zne,Wne\}}})\\[3pt]
\noindent zfeast\_g{\large \bf F}gv{\color{blue}{\{x\}}}({\bf \{List-A\}},{\bf \{List-B\}},fpm,epsout,loop,Emid,r,M0,E,X,M,res,info,{\color{blue}{\{Zne,Wne\}}})\\[3pt]
\noindent {\color{mygreen}{!!!! RCI (format independent) - Complex Symmetric and Real/Complex General}}\\[3pt]
\noindent zfeast\_srci{\color{blue}{\{x\}}}({\color{red}{ijob,N,Ze,work1,work2,Aq,Bq}},fpm,epsout,loop,Emid,r,M0,E,X,M,res,info,{\color{blue}{\{Zne,Wne\}}})\\[3pt]
\noindent zfeast\_grci{\color{blue}{\{x\}}}({\color{red}{ijob,N,Ze,work1,work2,Aq,Bq}},fpm,epsout,loop,Emid,r,M0,E,X,M,res,info,{\color{blue}{\{Zne,Wne\}}})\\[3pt]
\end{tt}

\noindent We note the following:
\begin{itemize}
\itemsep 1pt
\parskip 1pt
\item The Table below details the series of arguments in each {\tt \bf \{List-A\}}, and {\tt \bf \{List-B\}} that are specific to the type of matrix format represented above by
{\large \tt \bf F} (as a placeholder). 
\begin{center}
\begin{small}
\begin{tabular}{|c||c|c|c|}
\hline
\tt  & {\tt \bf F} & {\tt \bf List-A} & {\tt \bf List-B}  \\
\hline
\hline
\multicolumn{1}{|c||}{\col \tt Dense} & \col & \col & \col \\ \hline
Symmetric & \tt y &\tt  \{UPLO, N, A, LDA\} &\tt  \{B, LDB\} \\ \hline
General & \tt e &\tt  \{N, A, LDA\} &\tt  \{B, LDB\} \\ \hline
\multicolumn{1}{|c||}{\col \tt Banded} & \col & \col & \col \\ \hline
Symmetric & \tt b &\tt  \{UPLO, N, ka, A, LDA\} &\tt  \{kb, B, LDB\} \\ \hline
General & \tt b &\tt  \{N, kla, kua, A, LDA\} &\tt  \{klb, kub, B, LDB\} \\ \hline
\multicolumn{1}{|c||}{\col \tt Sparse} & \col & \col & \col \\ \hline
Symmetric & \tt csr &\tt  \{UPLO, N, A, IA, JA\} &\tt  \{B, IB, JB\} \\ \hline
General & \tt csr &\tt  \{N, A, IA, JA\} &\tt  \{B, IB, JB\} \\ \hline
\end{tabular}
\end{small}
\end{center}


\item  Similarly to the Hermitian case, Table \ref{tab-spec1} details the specific matrix-format
arguments in {\tt \bf \{List-A\}} and {\tt \bf \{List-B\}}.
For the banded drivers and the real/complex general cases,  {\tt kla} (resp.  {\tt klb}) represents
the number of sub-diagonals for matrix {\tt A} (resp. matrix {\tt B}), and  {\tt kua} (resp.  {\tt kub}) the number of super-diagonals for matrix {\tt A} (resp. matrix {\tt B}). 

\item Table \ref{tab-common2} details the common arguments in all the non-Hermitian FEAST interfaces above. 

\item Table \ref{tab-rci2} details the arguments for the non-Hermitian RCI interfaces (in red above).


\end{itemize}

\begin{table}[htbp]
\begin{center}
\begin{small}
\begin{tabular}{|l|p{5.5cm}|c|l|}
\hline
 & \multicolumn{1}{|c|}{Type} & I/O &\multicolumn{1}{|c|}{Description}  \\ 
\hline  \hline
{\tt ijob}  & 
integer  & in/out &  On entry: ijob=-1 (initialization)\\ 
& & & On exit:  ID of the FEAST\_RCI operation \\ \hline
{\tt N}  & integer  & in & Size of the system \\ \hline
{\tt Ze} & double complex   &  out    &  Coordinate along the complex contour  \\ \hline
{\tt work1} &   double complex({\tt N,M0})   &     in/out       
& Workspace               \\
&  {\em or}   &  & \\
& double complex({\tt N,2*M0})  & &  \\ 
& {\em (if left vector calculated for non-sym. interfaces and {\tt fpm(15)=0})}
& & \\ \hline
{\tt work2} &  double complex({\tt N,M0})   &  in/out    &  Workspace  \\  \hline
{\tt Aq, Bq} &  double complex({\tt M0,M0})  &     in/out       
& Workspace for the reduced eigenvalue problem         \\ \hline 
\end{tabular}
\caption{\label{tab-rci2} List of arguments for the FEAST RCI interfaces. Applicable to Non-Hermitian and Polynomial Drivers. }
\end{small}
\end{center}
\end{table}

\begin{table}[htb]
\begin{center}
\begin{small}
\begin{tabular}{|l|l|c|l|}
\hline
 & \multicolumn{1}{|c|}{Type} & I/O &\multicolumn{1}{|c|}{Description} \\ \hline \hline
{\tt fpm}  & 
integer(64)  & in/out & FEAST input parameters (see Table \ref{tab_fpm})\\ \hline
{\tt epsout}  & double real  & out & Trace relative error {\footnotesize $|{\tt trace}_k-{\tt trace}_{k-1}|/\max({\tt |Emid|+r})$}\\ \hline
{\tt loop} &  integer           &  out            &    \# of FEAST subspace iterations\\ \hline
{\tt Emid}  &   double complex   &  in  & Coordinate center of the contour ellipse \\ 
{\tt r}  &   double real   &  in  & Horizontal radius of the contour ellipse \\ \hline
{\tt M0}   &    integer         &    in/out          &  Search subspace dimension               \\ 
&                    &                          & On entry: initial guess {\tt M0$\geq$M} \\
&                    &                          & On exit:  new suitable {\tt M0} if guess too large    \\ \hline
{\tt E}    &  double complex({\tt M0})      &     in/out         & Eigenvalues    \\
&                              &                   &  On entry: initial guess if {\tt fpm(5)=1} (previous FEAST run) \\
&                              &                   &  On exit: Eigenvalues  solutions {\tt E(1:M)}   \\
           &                   &                 &  Remark: the {\tt E(M+1:M0)} values are outside the contour\\\hline 
{\tt X}    &   double complex({\tt N,M0})            &     in/out      & Eigenvectors ({\tt N}: size of the system)                \\
           &    {\em or}        &                & On entry: initial guess if {\tt fpm(5)=1} (previous FEAST run) \\
           &    double complex({\tt N,2*M0})           &                   & On exit: (right) Eigenvectors solutions  {\tt X(1:N,1:M)}\\
           &     {\em (if left vectors calculated for}                                &                       &Remarks: -left vectors (if calculated) in {\tt X(1:N,M0+1:M0+M)}  \\
& {\em non-sym. drivers and {\tt fpm(15)=0})}
&  & {\color{white}{Remarks:}} -if {\tt fpm(14)=1}, first $\bf Q$ subspace on exit  \\ \hline
{\tt M} &     integer        &     out         & \#Eigenvalues found inside contour   \\ 
 &             &             & \#Estimated eigenvalues if {\tt fpm(14)=2} \\  \hline
{\tt res} &  double complex({\tt M0})          &   out     & Relative residual {\tt res(1:M)} (right); {\tt res(M0+1,M0+M)} (left)  \\
      &    {\em or}   &        &   (right) $\bf ||Ax_i - \lambda_i Bx_i ||_2 /||\alpha Bx_i ||_2$  \\
&  double complex({\tt 2*M0}) & &  (left) $\bf ||A^Hx_i - \lambda_i^* B^Hx_i ||_2 /||\alpha B^Hx_i ||_2$ \\
&  {\em (if left vectors calculated)}  &   & with ${\tt \alpha}=\max({\tt |Emid|+r})$   \\ \hline
{\tt info} &  integer        &  out           &   Error handling  (see Table \ref{tab_info} for all INFO codes) \\ \hline
{\tt Zne,Wne}  &   double complex({\tt fpm(8)})    &  in  & Custom integration nodes and weights- Expert mode\\ \hline
\end{tabular}
\caption{\label{tab-common2} List of common arguments for the FEAST Driver interfaces. Applicable to Non-Hermitian and Polynomial Drivers.
}
\end{small}
\end{center}
\end{table}

\subsubsection*{\underline{non-Hermitian Driver Interfaces: Examples}}

Let us consider the following systems:

\begin{description}
\item[System3] a ``real non-symmetric'' generalized eigenvalue problem $\bf Ax=\lambda Bx$ , where  
$\bf A$ and $\bf B $ are real non-symmetric.
$\bf A$ and $\bf B$ are of the size $\rm N=1671$ and
have the same sparsity pattern with number of non-zero elements $\rm NNZ=13011$. 
\item[System4] a ``complex symmetric'' standard eigenvalue problem $\bf Ax=\lambda x$, where  
$\bf A$ is complex symmetric. 
$\bf A$ is of size $\rm N=801$ with number of non-zero elements $\rm NNZ=24591$.
\end{description}

The {\tt \$FEASTROOT/example/FEAST} directory provides Fortran and C implementation of these systems using both dense, banded and sparse-CSR storage. Here, the complete list of routines (System4 includes examples for using expert routines as well):

\begin{center}
\begin{tabular}{lll}
& \multicolumn{1}{c}{System3} &  \multicolumn{1}{c}{System4} \\ \hline\hline
\col dense & \col & \col \\
&{\tt \{F90,C\}dense\_dfeast\_gegv} &  {\tt \{F90,C\}dense\_zfeast\_syev\{x\}} \\
\col banded & \col & \col \\
&{\tt \{F90,C\}dense\_dfeast\_gbgv} &  {\tt \{F90,C\}dense\_zfeast\_sbev\{x\}} \\
\col sparse & \col & \col \\
&{\tt \{F90,C\}dense\_dfeast\_gcsrgv}  & {\tt \{F90,C\}dense\_zfeast\_scsrev\{x\}} \\
\hline
\end{tabular}
\end{center}

The {\tt \$FEASTROOT/example/PFEAST-L2} directory provides
the parallel implementation of all these routines using FEAST-MPI.
The routine names are preceded by the letter {\tt P}. 
The MPI parallelization operates only at the second level L2 where all the ({\tt fpm(2)}) linear systems are distributed among the MPI processes.

\subsubsection*{\underline{non-Hermitian RCI Interfaces}}

The RCI template for solving the complex symmetric problem is the same than the one used for solving the
real symmetric case in Section \ref{sec-Hermitian} (just replace {\tt dfeast\_srci\{x\}} by {\tt zfeast\_srci\{x\}}).

Here is a general (F90) template example of RCI for solving real/complex general problem:

\begin{small}
\begin{F90}%[frame=trBL]
ijob=-1 ! initialization
do while (ijob/=0)
call zfeast_grci(ijob,N,Ze,work1,work2,Aq,Bq,fpm,epsout,loop,Emid,r,M0,E,X,M,res,info)
 select case(ijob)
 case(10) !!Factorize the complex matrix Az <=(ZeB-A) - or factorize a preconditioner of ZeB-A
          !!REMARK:  Az can be formed and factorized using single precision arithmetic
................ <<< user entry
 case(11) !!Solve the linear system with fpm(23) rhs; Az * Qz=work2(1:N,1:fpm(23)) 
          !!Result (in place) in work2 <= Qz(1:N,1:fpm(23))
          !!REMARKS:  -Solve can be performed in single precision
          !!          -Low accuracy iterative solver are ok
................ <<< user entry
 case(20) !![Optional: *only if* needed by case(21)]
          !!Factorize the complex matrix Az^H
          !!REMARKS: -The matrix Az from case(10) cannot be overwritten  
          !!         -case(20) becomes obsolete if the solve in case(21) can be performed  
          !!           by reusing the factorization in case(10)
................ <<< user entry
 case(21) !!Solve the linear system with fpm(23) rhs;  Az^H * Qz=work2(1:N,1:fpm(23))
          !!Result (in place) in work2 <= Qz(1:N,1:fpm(23))          
................ <<< user entry
 case(30) !!Perform multiplication A * X(1:N,i:j) result in work1(1:N,i:j)
          !!         where i=fpm(24) and j=fpm(24)+fpm(25)-1
 ................ <<< user entry
 case(31) !!Perform multiplication A^H * X(1:N,i:j) result in work1(1:N,i:j)
          !!         where i=fpm(34) and j=fpm(34)+fpm(35)-1
................ <<< user entry
 case(40) !!Perform multiplication B * X(1:N,i:j) result in work1(1:N,i:j)
          !!        where i=fpm(24) and j=fpm(24)+fpm(25)-1
          !!REMARK: user must set work1(1:N,i:j)=X(1:N,i:j) if B=I
 ................ <<< user entry
 case(41) !!Perform multiplication B^H * X(1:N,i:j) result in work1(1:N,i:j)
          !!        where i=fpm(34) and j=fpm(34)+fpm(35)-1
          !!REMARK: user must set work1(1:N,i:j)=X(1:N,i:j) if B=I
................  <<< user entry
 end select
end do
 \end{F90}
\end{small}

\newpage

\subsection{FEAST Polynomial (quadratic, cubic, quartic, etc.)}
Solving  $\displaystyle\sum_{i=0}^p \lambda^iA_ix=0$

\begin{tt}
\noindent {\color{mygreen}{!!!!  \{Ai\} Real-Sym., Complex Herm., Complex Sym., Real General and Complex General }}\\[3pt]
\noindent dfeast\_s{\large \bf F}pev{\color{blue}{\{x\}}}({\bf \{List-A\}},fpm,epsout,loop,Emid,r,M0,E,X,M,res,info,{\color{blue}{\{Zne,Wne\}}})\\[3pt]
\noindent zfeast\_h{\large \bf F}pev{\color{blue}{\{x\}}}({\bf \{List-A\}},fpm,epsout,loop,Emid,r,M0,E,X,M,res,info,{\color{blue}{\{Zne,Wne\}}}) \\[3pt]
\noindent zfeast\_s{\large \bf F}pev{\color{blue}{\{x\}}}({\bf \{List-A\}},fpm,epsout,loop,Emid,r,M0,E,X,M,res,info,{\color{blue}{\{Zne,Wne\}}}) \\[3pt]
\noindent dfeast\_g{\large \bf F}pev{\color{blue}{\{x\}}}({\bf \{List-A\}},fpm,epsout,loop,Emid,r,M0,E,X,M,res,info,{\color{blue}{\{Zne,Wne\}}}) \\[3pt]
\noindent zfeast\_g{\large \bf F}pev{\color{blue}{\{x\}}}({\bf \{List-A\}},fpm,epsout,loop,Emid,r,M0,E,X,M,res,info,{\color{blue}{\{Zne,Wne\}}}) \\[3pt]
\noindent {\color{mygreen}{!!!! RCI (format independent) - Real/Complex Symmetric and Hermitian/Real/Complex General}}\\[3pt]
\noindent zfeast\_srcipev{\color{blue}{\{x\}}}({\color{red}{ijob}},{\color{orange}{p}},{\color{red}{N,Ze,work1,work2,Aq,Bq}},fpm,epsout,loop,Emid,r,M0,E,X,M,res,info,{\color{blue}{\{Zne,Wne\}}})\\[3pt]
\noindent zfeast\_grcipev{\color{blue}{\{x\}}}({\color{red}{ijob}},{\color{orange}{p}},{\color{red}{N,Ze,work1,work2,Aq,Bq}},fpm,epsout,loop,Emid,r,M0,E,X,M,res,info,{\color{blue}{\{Zne,Wne\}}})\\[3pt]
\end{tt}

\noindent We note the following:
\begin{itemize}
\itemsep 1pt
\parskip 1pt
\item The Table below details the series of arguments in {\tt \bf \{List-A\}} that are specific to the type of matrix format represented above by {\large \tt \bf F} (as a placeholder). Remark: the banded format is not supported.
\begin{center}
\begin{small}
\begin{tabular}{|c||c|c|}
\hline
\tt  & {\tt \bf F} & {\tt \bf List-A}  \\
\hline
\hline
\multicolumn{1}{|c||}{\col \tt Dense} & \col & \col  \\ \hline
Symmetric/Hermitian & \tt \{y,e\} &\tt  \{UPLO, p, N, A, LDA\}  \\ \hline
General & \tt e &\tt  \{p, N, A, LDA\}  \\ \hline
\multicolumn{1}{|c||}{\col \tt Sparse} & \col & \col \\ \hline
Symmetric/Hermitian & \tt csr &\tt  \{UPLO, p, N, A, IA, JA\}  \\ \hline
General & \tt csr &\tt  \{p, N, A, IA, JA\}  \\ \hline
\end{tabular}
\end{small}
\end{center}

\item  Table \ref{tab-spec3} details the specific matrix-format arguments in {\tt \bf \{List-A\}}.

\item  The common arguments in all the polynomial FEAST interfaces above are identical to the ones given for the non-Hermitian case in Table \ref{tab-common2}.

\item The argument for the RCI interfaces (in red above) are also identical to the ones given for the non-Hermitian
case in Table \ref{tab-common2}. We note the addition of the argument integer {\color{orange}{\tt p}} which stands for the degree
of the polynomial (e.g. {\tt p=2} fro quadratic,  {\tt p=3} for cubic etc.) 

\end{itemize}

\begin{table}[htbp]
\begin{center}
\begin{small}
\begin{tabular}{|l|p{6.3cm}|c|l|}
\hline
 & \multicolumn{1}{|c|}{Type} & I/O &\multicolumn{1}{|c|}{Description}  \\ 
\hline  \hline
   \multicolumn{4}{|l|}{\col Common}    \\ \hline
{\tt UPLO}  & 
character(len=1) & in & Matrix Storage {\tt ('F','L','U')} \\ 
& & & 'F': Full; 'L': Lower; 'U': Upper\\ \hline
{\tt N}  & integer  & in & Size of the system \\ \hline
{\tt p}  & integer  & in & Degree of the polynomial \\ \hline
\multicolumn{4}{|l|}{\col Dense}   \\ \hline
{\tt A}  &  double real({\tt LDA,N,p+1}) \hfill using {\tt dfeast}    &     in       
& All system matrices {\tt A(:,i)}        \\
  &  double complex({\tt LDA,N,p+1})  \hfill using {\tt zfeast}    &           
&         \\ \hline
{\tt LDA}  &    integer          &     in       
& 1st Leading dimension of {\tt A} {\tt LDA>=N}  \\ \hline
\multicolumn{4}{|l|}{\col Sparse-csr - where {\tt nnz\_max} stands for the max of non-zero elements among all A sparse matrices}
  \\ \hline
{\tt A}  &  double real({\tt nnz\_max,p+1})  \hfill using {\tt dfeast}    &     in       
&  All system matrices {\tt A(:,i)} - CSR values         \\
  &  double complex({\tt nnz\_max,p+1})  \hfill using {\tt zfeast}    &           
&          \\ \hline
{\tt IA}  &   integer({\tt N+1,p+1})          &     in       
  & All sparse CSR Row array of {\tt A(:,i)}.          \\ \hline
{\tt JA}  &   integer({\tt nnz\_max,p+1})          &     in       
& All sparse CSR Column array of {\tt A(:,i)}.          \\ \hline
\end{tabular}
\end{small}
\caption{\label{tab-spec3}   List of arguments that are matrix-format specific for the FEAST Polynomial  Driver interfaces.
 Remark:  {\tt A(:,i)}==$A_{i-1}$ in Fortran.}
\end{center}
\end{table}

\subsubsection*{\underline{Polynomial Driver Interfaces: Examples}}

Let us consider the following system:

\begin{description}
\item[System5] a quadratic eigenvalue problem $\bf (A_2\lambda^2+A_1\lambda+A_0)x=0$ , where  
$\bf A_2,A_1,A_0$ are  real symmetric. The size of the system is $\rm N=1000$ with
  $2998$ non-zero elements for $A_0$ and $A_1$, and $1000$ for $A_2$. 
\end{description}

The {\tt \$FEASTROOT/example/FEAST} directory provides Fortran and C implementation of this system using both dense and sparse-CSR storage. Here, the complete list of routines:

\begin{center}
\begin{tabular}{ll}
& \multicolumn{1}{c}{System5} \\ \hline\hline
\col dense & \col\\
&{\tt \{F90,C\}dense\_dfeast\_sypev}  \\
\col sparse & \col \\
&{\tt \{F90,C\}dense\_dfeast\_scsrpev}   \\
\hline
\end{tabular}
\end{center}

The {\tt \$FEASTROOT/example/PFEAST-L2} directory provides
the parallel implementation of these routines using FEAST-MPI.
The routine names are preceded by the letter {\tt P}. 
The MPI parallelization operates only at the second level L2 where all the ({\tt fpm(2)}) linear systems are distributed among the MPI processes.

\subsubsection*{\underline{Polynomial RCI Interfaces}}

Here is a general (F90) template example of RCI for solving real/complex symmetric problem:

\begin{small}
  \begin{F90}%[frame=trBL]
!!! Here your polynomial matrices are in stored A[i] (i=1,..,p+1) (p polynomial degree)
!!! All the matrices are real or complex symmetric
ijob=-1 ! initialization
do while (ijob/=0)
call zfeast_srcipev(ijob,p,N,Ze,work1,work2,Aq,Bq,fpm,epsout,loop,Emid,r,M0,E,X,M,res,info)
 select case(ijob)
 case(10) !!Form and Factorize P(Ze)
          !!Example for the quadratic problem: P(Ze)=A[3]*Ze**2+A[2]*Ze+A[1]
          !!REMARK:  P(Ze) can be formed and factorized using single precision arithmetic
................ <<< user entry
 case(11) !!Solve the linear system with fpm(23) rhs; P(Ze)* Qz=work2(1:N,1:fpm(23)) 
          !!Result (in place) in work2 <= Qz(1:N,1:fpm(23))
          !!REMARKS:  -Solve can be performed in single precision
          !!          -Low accuracy iterative solver are ok
................ <<< user entry
 case(30) !!Perform multiplication A[fpm(57)] * X(1:N,i:j) result in work1(1:N,i:j)
          !!         where i=fpm(24) and j=fpm(24)+fpm(25)-1; fpm(57) take the values 1...p+1
 ................ <<< user entry
 end select
end do
 \end{F90}
\end{small}

Here is a general (F90) template example of RCI for solving Hermitian or real/complex general problem:

\begin{small}
  \begin{F90}%[frame=trBL]
!!! Here your polynomial matrices are in stored A[i] (i=1,..,p+1) (p polynomial degree)
!!! At least one matrix is not real/complex symmetric
ijob=-1 ! initialization
do while (ijob/=0)
call zfeast_grcipev(ijob,p,N,Ze,work1,work2,Aq,Bq,fpm,epsout,loop,Emid,r,M0,E,X,M,res,info)
 select case(ijob)
 case(10) !!Form and Factorize P(Ze)
          !!Example for the quadratic problem: P(Ze)=A[3]*Ze**2+A[2]*Ze+A[1]
          !!REMARK:  P(Ze) can be formed and factorized using single precision arithmetic
................ <<< user entry
 case(11) !!Solve the linear system with fpm(23) rhs; P(Ze)* Qz=work2(1:N,1:fpm(23)) 
          !!Result (in place) in work2 <= Qz(1:N,1:fpm(23))
          !!REMARKS:  -Solve can be performed in single precision
          !!          -Low accuracy iterative solver are ok
................ <<< user entry
 case(20) !![Optional: *only if* needed by case(21)]
          !!Factorize the complex matrix P(Ze)^H
          !!REMARKS: -The factorization P(Ze) from case(10) cannot be overwritten  
          !!         -case(20) becomes obsolete if the solve in case(21) can be performed  
          !!           by reusing the factorization in case(10)
................ <<< user entry
 case(21) !!Solve the linear system with fpm(23) rhs;  P(Ze)^H * Qz=work2(1:N,1:fpm(23))
          !!Result (in place) in work2 <= Qz(1:N,1:fpm(23))          
................ <<< user entry
 case(30) !!Perform multiplication A[fpm(57)] * X(1:N,i:j) result in work1(1:N,i:j)
          !!         where i=fpm(24) and j=fpm(24)+fpm(25)-1; fpm(57) will the values 1...p+1
 ................ <<< user entry
 case(31) !!Perform multiplication A^H[fpm(57)] * X(1:N,i:j) result in work1(1:N,i:j)
          !!         where i=fpm(34) and j=fpm(34)+fpm(35)-1; fpm(57) will the values 1...p+1
................ <<< user entry
 end select
end do
 \end{F90}
\end{small}

\newpage

\subsection{IFEAST (FEAST w/o Factorization)}

IFEAST stands for using FEAST using inexact iterative solver for solving the linear systems (instead of using a direct solver).
IFEAST only supports the sparse driver interfaces where the MKL-PARDISO solver is then replaced by a built-in BiCGstab solver.
IFEAST is particularly effective if the sparse system matrix is very large (typically >1M) and/or the direct factorization
becomes too expensive (both in time and memory).
Two options are possible for calling IFEAST:
\begin{description}
\itemsep 1pt
\parskip 1pt
\item[Option1] In the naming convention of all FEAST sparse drivers, replace {\tt feast} by {\tt ifeast}.  
\item[Option2] Keep the name of your FEAST sparse driver unchanged but use the new flag value {\tt fpm(43)=1}. 
\end{description}

As an example, let us re-work the helloworld example presented in Section \ref{sec-hello}.
Here are the few lines that need to be changed in the code in order to make use of IFEAST:

\begin{small}
\begin{F90}
!!!! the matrix A needs first to be defined in sparse CSR format;
!!!! add these lines in the variable declaration section of the program
integer, parameter :: NNZ
double precision,dimension(NNZ) :: sA=(/2.0d0,-1.0d0,-1.0d0,-1.0d0,3.0d0,-1.0d0,-1.0d0,&
                                     &-1.0d0,-1.0d0,3.0d0,-1.0d0,-1.0d0,-1.0d0,2.0d0/)
integer,dimension(N+1) :: IA=(/1,4,8,12,15/)
integer,dimension(NNZ) :: JA=(/1,2,3,1,2,3,4,1,2,3,4,2,3,4/)

    
!!!! The direct call to IFEAST can be done as follow (option 1)
call feastinit(fpm)
fpm(1)=1 !! change from default value (print info on screen)
call difeast_scsrev(UPLO,N,sA,IA,JA,fpm,epsout,loop,Emin,Emax,M0,E,X,M,res,info)

!!!! Alternatively, an indirect call to IFEAST is also possible (option 2)
call feastinit(fpm)
fpm(1)=1 !! change from default value (print info on screen)
fpm(43)=1 !! switch solver in FEAST sparse driver from MKL-PARDISO to BicGStab
call dfeast_scsrev(UPLO,N,sA,IA,JA,fpm,epsout,loop,Emin,Emax,M0,E,X,M,res,info)
\end{F90}
\end{small}    

Here is the output of the run:

\begin{shaded}
\begin{footnotesize}
\begin{Verbatim}[frame=single,baselinestretch=0.9]
***********************************************
*********** FEAST v4.0 BEGIN ******************
***********************************************
Routine DIFEAST_SCSREV
Solving AX=eX with A real symmetric
List of input parameters fpm(1:64)-- if different from default
   fpm( 1)=   1
 
.-------------------.
| FEAST data        |
--------------------.-------------------------.
| Emin              |  3.0000000000000000E+00 |
| Emax              |  5.0000000000000000E+00 |
| #Contour nodes    |  4   (half-contour)     |
| Quadrature rule   |  Trapezoidal            |
| Ellipse ratio y/x |  1.00                   |
| System solver     |  BiCGstab               |
|                   |  eps=1E-1; maxit=  40   |
|                   |  Single precision       |
|                   |  Matrix scaled          |
| FEAST uses MKL?   |  Yes                    |
| Fact. stored?     |  Yes                    |
| Initial Guess     |  Random                 |
| Size system       |      4                  |
| Size subspace     |      3                  |
-----------------------------------------------
 
.-------------------.
| FEAST runs        |
----------------------------------------------------------------------------------------------
#It |  #Eig  |          Trace            |     Error-Trace          |     Max-Residual
----------------------------------------------------------------------------------------------
  #it     4; res min=   1.0501874385226984E-05; res max=   3.1005099299363792E-04
  #it     3; res min=   1.8933478742837906E-02; res max=   9.8125219345092773E-02
  #it     2; res min=   2.9427373781800270E-02; res max=   8.7035216391086578E-02
  #it     2; res min=   1.8594998866319656E-02; res max=   4.0714181959629059E-02
  0      2      7.9995148090616173E+00      1.0000000000000000E+00      8.7043125405406700E-03
  #it     2; res min=   2.8199069201946259E-02; res max=   5.2359659224748611E-02
  #it     2; res min=   6.4112566411495209E-02; res max=   8.7011836469173431E-02
  #it     2; res min=   3.0766267329454422E-02; res max=   5.8814667165279388E-02
  #it     2; res min=   1.4842565171420574E-02; res max=   1.5974638983607292E-02
  1      2      7.9999999809553799E+00      9.7034378752525186E-05      5.5062636929826442E-05
  #it     1; res min=   5.3539618849754333E-02; res max=   5.3563684225082397E-02
  #it     1; res min=   8.7251774966716766E-02; res max=   8.7261073291301727E-02
  #it     1; res min=   7.2112381458282471E-02; res max=   7.2119705379009247E-02
  #it     1; res min=   6.5362729132175446E-02; res max=   6.5378636121749878E-02
  2      2      7.9999999999999591E+00      3.8089158493903597E-09      7.6857681986636782E-08
  #it     1; res min=   3.4910961985588074E-02; res max=   3.4911289811134338E-02
  #it     1; res min=   8.4501713514328003E-02; res max=   8.4501914680004120E-02
  #it     1; res min=   6.9185368716716766E-02; res max=   6.9185607135295868E-02
  #it     1; res min=   6.2304046005010605E-02; res max=   6.2304504215717316E-02
  3      2      8.0000000000000036E+00      8.8817841970012523E-15      2.5198258072819870E-12
  #it     1; res min=   3.4680232405662537E-02; res max=   3.4876041114330292E-02
  #it     1; res min=   8.4406383335590363E-02; res max=   8.4497839212417603E-02
  #it     1; res min=   6.9181196391582489E-02; res max=   6.9283291697502136E-02
  #it     1; res min=   6.2299706041812897E-02; res max=   6.2503643333911896E-02
  4      2      8.0000000000000000E+00      7.1054273576010023E-16      3.0767402982137245E-16
 
==>FEAST has successfully converged with Residual tolerance <1E-12
   # FEAST outside it.        4
   # Inner BiCGstab it.      31
   # Eigenvalue found         2 from   3.9999999999999991E+00 to   4.0000000000000009E+00
----------------------------------------------------------------------------------------------
 
.-------------------.
| FEAST-RCI timing  |
--------------------.------------------.
| Fact. cases(10,20)|      0.0000      |
| Solve cases(11,12)|      0.0189      |
| A*x   cases(30,31)|      0.0000      |
| B*x   cases(40,41)|      0.0001      |
| Misc. time        |      0.0005      |
| Total time (s)    |      0.0195      |
--------------------------------------- 
 
***********************************************
*********** FEAST- END*************************
***********************************************
 
 Solutions (Eigenvalues/Eigenvectors/Residuals)
 E=   4.00000000000000      X=  0.353553390593291      -0.853553390593266     
  0.146446609406686       0.353553390593291      Res=  3.076740298213724E-016
 
 E=   4.00000000000000      X=  0.353553390593257       0.146446609406767     
 -0.853553390593281       0.353553390593257      Res=  2.361858070262224E-016
\end{Verbatim}
\end{footnotesize}
\end{shaded}

{\bf Some Remarks:}
\begin{itemize}
\itemsep 1pt
\parskip 1pt
\item IFEAST is using different default {\tt fpm} parameters than FEAST. In particular, the trapezoidal rule is used along a half-circle with 4 contour integration points, as well as the BiCGStab iterative solver.
\item Each FEAST loop reports the total number of BiCGstab iterations for each linear system solve needed to reach the accuracy defined in {\tt fpm(45)}. The total number of BiCGtab iterations will be reported in {\tt fpm(60)} (here 31, of course IFEAST is rather
ineffective for such small system).
\item You may have noticed that the values of the output eigenvectors for this example are different than the ones reported in Section \ref{sec-hello}. As a reminder, eigenvectors are not unique, both solutions are here orthonormal and correct (they span the same eigenvector subspace).
\end{itemize}

\newpage

\subsection{PFEAST and PIFEAST (MPI-solver)}\label{sec-pfeast}



In FEAST v4.0, the FEAST-MPI library enables the use of MPI linear system solvers at L3.
As a result, the three level of parallelisms of FEAST (L1-L2-L3) can all support MPI (FEAST is internally using three MPI communicators).
This MPI-MPI-MPI programming model is named PFEAST.
PFEAST currently supports all the sparse drivers (for Hermitian/non-Hermitian/Polynomial problems)  as well as all the RCI interfaces.
Using RCI, an expert developer could straightforwardly customize PFEAST using
 highly-efficient application-specific MPI solvers such as: domain decompositions, or iterative/hybrid solvers with/without preconditioners.

Here some information about the use of PFEAST:
\begin{description}
\item[Initialization] The {\tt feastinit(fpm)} routine must be replaced by \fbox{\tt pfeastinit(fpm, L1\_comm\_world, nL3)} which, in addition of setting up default {\tt fpm} values,
is going to initialize all MPI communicators.
\begin{itemize}
\itemsep 1pt
\parskip 1pt
\item {\tt L1\_comm\_world} represents your own defined MPI communicator for a given search contour (containing {\tt nL1} total MPI processes), 
if only a single contour is used it must take the value {\tt MPI\_COMM\_WORLD}.
\item {\tt nL3} is an (in/out) integer input that indicates the number of MPI processes you wish to use at level L3 (MPI system solver). The value of {\tt nL3} will be reassigned to {\tt nL1} if {\tt nL1} is not a multiple of {\tt nL3}. In addition, if  {\tt nL1}< {\tt nL3} then {\tt nL3}= {\tt nL1} on exit.  Ideally, {\tt nL1}/{\tt nL3} should be a multiple of a number of contour points (if it is equal to the number of contour points, then L2 is optimally used). Furthermore, L3 can also be threaded (MPI calling OpenMP on each local distributed system), make sure that your number of selected threads {\tt <omp>} times {\tt nL1}  does not exceed the number of available physical cores of your cluster.

\item This initialization routine is setting up the L3 communicator {\tt fpm(49)} (n particular) that may be needed to distribute your matrix.
\end{itemize}
\item[Sparse Drivers] In the naming convention of the FEAST sparse drivers, all routine names must be preceded by the letter {\tt p}.
In particular, you must replace {\tt z\{i\}feast} by {\tt pz\{i\}feast}, or {\tt d\{i\}feast} by {\tt pd\{i\}feast}.
We actually use the names PFEAST and PIFEAST to indicate the MPI version of the FEAST and IFEAST interfaces, respectively.
PFEAST is using MKL-Cluster-PARDISO and PIFEAST is using a built-in MPI-BiCGstab iterative solver (PBiCGStab). PIFEAST is also using
its own built-in highly efficient MPI sparse mat-vec library.

\noindent PFEAST/PIFEAST drivers allow two  options for the row distribution of matrices and solution vectors:
\begin{itemize}
\itemsep 1pt
\parskip 1pt
\item  global and common to all L3-MPI processes (The row-distribution will then take place internally).
The argument list for all interfaces stay unchanged. 
\item locally row distributed among all L3-MPI processes. The user is responsible for distributing the data using the {\tt fpm(49)}
L3 communicator. The argument list for all interfaces stay mostly unchanged but the size of matrix/vectors {\tt N} which must now be local. 
\end{itemize}
Furthermore, PFEAST/PIFEAST will automatically detect which option above you are using!
\item[RCI Interfaces] The names of the RCI interfaces do not change (the letter {\tt p} is not needed). In the argument list, only
the size {\tt N}  must be changed to its local value (i.e. local number of rows for the row-distributed vector and work arrays).
\end{description}

\subsubsection*{\underline{PFEAST with global L3 distribution: Examples}}

The {\tt \$FEASTROOT/example/PFEAST-L2L3} directory provides
Fortran and C implementation of System1 to System5 examples (discussed previously). Here is the complete list of the PFEAST routines that are all using global sparse CSR-storage.

\begin{center}
\begin{tabular}{lll}
 & Type & Routine \\ \hline\hline
\col System1 & \col Real Sym. Generalized  & \col {\tt P\{F90,C\}sparse\_pdfeast\_scsrgv} \\
 System2 & Complex Herm. Standard &{\tt P\{F90,C\}sparse\_pzfeast\_hcsrev}\\
\col System3 & \col Real non-Sym. Generalized & \col {\tt P\{F90,C\}sparse\_pdfeast\_gcsrgv}\\
System4 & Complex Sym. Standard &{\tt P\{F90,C\}sparse\_pzfeast\_scsrev}\\
\col System5 & \col Real Sym. Quadratic &\col {\tt P\{F90,C\}sparse\_pdfeast\_scsrpev}\\ \hline
\end{tabular}
\end{center}

\subsubsection*{\underline{PFEAST with local L3 distribution: Example}}

As an example, let us re-work the helloworld example presented in Section \ref{sec-hello} using 2 MPI processes to distributed the $4\times 4$ matrix. We obtain (for example):

$$
{\bf A}=
\begin{pmatrix}
 2 & -1 & -1 & 0\\
 -1 & 3 & -1 &-1\\ \hline
 -1 &-1&3 &-1 \\
 0 &-1 & -1 & 2
\end{pmatrix}
=\begin{pmatrix}
{\bf A_1} \\ \hline
{\bf A_2}
\end{pmatrix}
$$

\noindent A Fortran90 source code example is provided below:

\begin{small}
  \begin{F90}%[frame=trBL]
program helloworld_pfeast_local
  implicit none
  include 'mpif.h'
  !! 4x4 global eigenvalue system == two 2x4 local matrices 
  integer,parameter :: Nloc=2, NNZloc=7
  character(len=1) :: UPLO='F'
  double precision,dimension(NNZloc) ::  Aloc
  integer,dimension(Nloc+1) :: IAloc
  integer,dimension(NNZloc) :: JAloc
  !! input parameters for FEAST
  integer,dimension(64) :: fpm
  integer :: M0=3 ! search subspace dimension
  double precision :: Emin=3.0d0, Emax=5.0d0 ! search interval
  !! output variables for FEAST
  double precision,dimension(:),allocatable :: E, res
  double precision,dimension(:,:),allocatable :: X
  double precision :: epsout
  integer :: nL3,rank3,loop,info,M,i
!!! MPI
  integer :: code
  call MPI_INIT(code)

!!! Allocate memory for eigenvalues.eigenvectors,residual
  allocate(E(M0),X(Nloc,M0),res(M0))

!!!!!!!!!! INITIALIZE PFEAST and DISTRIBUTE MATRIX
  nL3=2
  call pfeastinit(fpm,MPI_COMM_WORLD,nL3)

  call MPI_COMM_RANK(fpm(49),rank3,code) !! find rank of new L3 communicator
  if (rank3==0) then
    Aloc=(/2.0d0,-1.0d0,-1.0d0,-1.0d0,3.0d0,-1.0d0,-1.0d0/)
    IAloc=(/1,4,8/)
    JAloc=(/1,2,3,1,2,3,4/)
  elseif (rank3==1) then
    Aloc=(/-1.0d0,-1.0d0,3.0d0,-1.0d0,-1.0d0,-1.0d0,2.0d0/)
    IAloc=(/1,5,8/)
    JAloc=(/1,2,3,4,2,3,4/)
  endif
    
!!!!!!!!!! PFEAST
  fpm(1)=1 !! change from default value (print info on screen)  
  call pdfeast_scsrev(UPLO,Nloc,Aloc,IAloc,JAloc,fpm,epsout,loop,Emin,Emax,M0,E,X,M,res,info)

!!!!!!!!!! REPORT
 if (info==0) then
     print *,'Solutions (Eigenvalues/Eigenvectors/Residuals) at rank L3',rank3
     do i=1,M
        print *,'Eigenvalue',i
        print *,'E=',E(i),'X=',X(:,i),'Res=',res(i)
     enddo
  endif

end program helloworld_pfeast_local
\end{F90}
\end{small}

\noindent  Your program must be compiled using the same MPI implementation used to compile the FEAST-MPI library.
Once compiled, your source program must now be linked with the {\tt pfeast} library. You can use (for example):
\begin{itemize}
\item  \fbox{\parbox{\dimexpr\linewidth-2\fboxsep-2\fboxrule\relax}{\tt mpiifort -o helloworld\_pfeast\_local helloworld\_pfeast\_local.f90  -L\$FEASTROOT>/lib/<{\it arch}>  -lpfeast -mkl=parallel -lmkl\_blacs\_intelmpi\_lp64 -liomp5 -lpthread -lm -ldl}} \\
  if FEAST was compiled with {\tt ifort}, MKL flag was set to 'yes', and MPI was chosen to be 'impi' (intel mpi).
\item \fbox{\parbox{\dimexpr\linewidth-2\fboxsep-2\fboxrule\relax}{\tt mpif90.mpich -fc=gfortran -o helloworld\_pfeast\_local helloworld\_pfeast\_local.f90  -L\$FEASTROOT>/lib/<{\it arch}>  -lpfeast -Wl,--no-as-needed -lmkl\_gf\_lp64 -lmkl\_gnu\_thread -lmkl\_core -lmkl\_blacs\_intelmpi\_lp64 -lgomp -lpthread -lm -ldl}}\\
   if FEAST was compiled with {\tt gfortran}, MKL flag was set to 'yes', and MPI was chosen to be 'mpich'.
\end{itemize}

\noindent A run of the resulting executable looks like:
         \begin{center}\fbox{\tt mpirun -n 2 ./helloworld\_pfeast\_local} \end{center}
and the output of the run should be:

\begin{shaded}
\begin{footnotesize}
\begin{Verbatim}[frame=single,baselinestretch=0.9]
***********************************************
*********** FEAST v4.0 BEGIN ******************
***********************************************
Routine PDFEAST_SCSREV
Solving AX=eX with A real symmetric
#MPI (total=L2*L3)    2=   1*   2
List of input parameters fpm(1:64)-- if different from default
   fpm( 1)=   1
 
.-------------------.
| FEAST data        |
--------------------.-------------------------.
| Emin              |  3.0000000000000000E+00 |
| Emax              |  5.0000000000000000E+00 |
| #Contour nodes    |  8   (half-contour)     |
| Quadrature rule   |  Gauss                  |
| Ellipse ratio y/x |  0.30                   |
| System solver     |  MKL-Cluster-Pardiso    |
|                   |  Single precision       |
|                   |  Matrix scaled          |
| FEAST uses MKL?   |  Yes                    |
| Fact. stored?     |  Yes                    |
| Initial Guess     |  Random                 |
| Size system       |      4                  |
| Size subspace     |      3                  |
-----------------------------------------------
 
.-------------------.
| FEAST runs        |
----------------------------------------------------------------------------------------------
#It |  #Eig  |          Trace            |     Error-Trace          |     Max-Residual
----------------------------------------------------------------------------------------------
  0      2      7.9999999999999947E+00      1.0000000000000000E+00      1.2829791863202860E-08
  1      2      8.0000000000000000E+00      1.0658141036401502E-15      6.6022321739723205E-16
 
==>FEAST has successfully converged with Residual tolerance <1E-12
   # FEAST outside it.        1
   # Eigenvalue found         2 from   3.9999999999999996E+00 to   4.0000000000000000E+00
----------------------------------------------------------------------------------------------
 
.-------------------.
| FEAST-RCI timing  |
--------------------.------------------.
| Fact. cases(10,20)|      0.0058      |
| Solve cases(11,12)|      0.0025      |
| A*x   cases(30,31)|      0.0000      |
| B*x   cases(40,41)|      0.0000      |
| Misc. time        |      0.0005      |
| Total time (s)    |      0.0088      |
--------------------------------------- 
 
***********************************************
*********** FEAST- END*************************
***********************************************
 
 Solutions (Eigenvalues/Eigenvectors/Residuals) at rank L3           0
 Eigenvalue           1
 E=   4.00000000000000      X=  0.358971274554087      -0.851189974005894      Res=  2.784560286672102E-016
 Eigenvalue           2
 E=   4.00000000000000      X= -0.348051180209196      -0.159610864767551      Res=  6.602232173972321E-016
 Solutions (Eigenvalues/Eigenvectors/Residuals) at rank L3           1
 Eigenvalue           1
 E=   4.00000000000000      X=  0.133247424897719       0.358971274554087      Res=  2.784560286672102E-016
 Eigenvalue           2
 E=   4.00000000000000      X=  0.855713225185942      -0.348051180209196      Res=  6.602232173972321E-016
\end{Verbatim}
\end{footnotesize}
\end{shaded}

\subsubsection*{\underline{PFEAST using 3 levels of parallelism: Example}}

Three levels of parallelism  means that L1 is active and
multiple search contours can be used simultaneously.
FEAST v4.0 does not offer automatic partitioning of the overall eigenvalue spectrum, it is then up to the users to guess it.
Users could take advantage
of  fast stochastic estimates with the flag {\tt fpm(14)=2}.
Once the eigenvalue spectrum partitioned, a single
call to PFEAST will account for all L1-L2-L3 MPI parallelism.

The {\tt \$FEASTROOT/example/PFEAST-L1L2L3} directory provides
Fortran and C implementation of the System2 example (discussed previously). It uses two search intervals. The name of the routines are:


\begin{center}
\begin{tabular}{ll}
&  \multicolumn{1}{c}{System2} \\ \hline\hline
\col sparse  & 
 {\tt 3P\{F90,C\}dense\_pzfeast\_hcsrev} \\
\hline
\end{tabular}
\end{center}

\noindent {\bf Remark:} Since multiple search intervals are involved, the option {\tt fpm(1)=1} (printing FEAST info on screen), may
provide a bit confusing results to read. You can easily change this flag value using {\tt fpm(1)=-i}
with {\tt i} is associated with the {\tt rank i-1} of the L1 MPI Communicator. Each search contour will then print all
its FEAST results into separate files named {\tt feast\{i\}.log}.

\newpage


\newpage

\section{Complement}
\subsection{Matrix storage}\label{sec_format}

Let us consider the following  matrix {\bf A} (as an example): \\[5pt]

\begin{equation}
{\bf A}=
\begin{pmatrix}
 a_{11} & a_{12} & 0 & 0\\
a_{21} & a_{22} & a_{23}& 0  \\
0 & a_{32} & a_{33} & a_{34} \\
0 & 0 & a_{43} & a_{44} \\  
 \end{pmatrix}
\end{equation}\\[5pt]
If the matrix presents some particular properties
such as Hermitian ($a_{ij}=a_{ji}^*$ for $i\neq j$) or symmetric ($a_{ij}=a_{ji}$),
only half of  the matrix elements need to be defined.   
Using the FEAST Driver interfaces, this matrix could be stored in dense, banded or sparse-CSR format as follows:
\begin{description}

\item[Dense] The matrix is stored in a two dimensional array in a straightforward fashion. 
Using the options {\tt UPLO='L'} or {\tt UPLO='U'}, the lower triangular or upper triangular part respectively,
do not need to be referenced.

\item[Banded] The matrix is also stored in a two dimensional array  following the 
banded LAPACK-type storage:\\[5pt]
$$
{\bf A=}
\begin{pmatrix}
* & a_{12} & a_{23}& a_{34}  \\
a_{11} & a_{22} & a_{33}& a_{44}  \\
a_{21} & a_{32} & a_{43} & * 
 \end{pmatrix}
$$\\[5pt]
In contrast to LAPACK, no extra-storage space is necessary since {\tt LDA>=kl+ku+1} if {\tt UPLO='F'} (LAPACK banded storage 
would require {\tt LDA>=2*kl+ku+1}). For this example, the number of subdiagonals {\tt kl} and superdiagonals is {\tt ku} are both equal to 1. 
Using the option  {\tt UPLO='L'} or {\tt UPLO='U'}, the  rows respectively 
above or below the diagonal elements row, do not need to be referenced (or stored).

\item[Sparse-CSR] The non-zero elements of the matrix are stored using a set of one 
dimensional arrays ({\tt A,IA,JA})  following the definition of the CSR (Compressed Sparse Row) format
$$
\begin{array}{rl}
{\bf A=}&(a_{11},a_{12},a_{21},a_{22},a_{23},a_{32},a_{33},a_{34},a_{43},a_{44}) \\ 
{\bf IA=}&(1,3,6,9,11) \\
{\bf JA=}&(1,2,1,2,3,2,3,4,3,4) 
\end{array}
$$
Using the option {\tt UPLO='L'} or {\tt UPLO='U'}, one would get respectively

$$
\begin{array}{rl}
{\bf A=}&(a_{11},a_{21},a_{22},a_{32},a_{33},a_{43},a_{44}) \\ 
{\bf IA=}&(1,2,4,6,8) \\
{\bf JA=}&(1,1,2,2,3,3,4) 
\end{array}
\mbox{~~~~and~~~~} \hfill
\begin{array}{rl}
{\bf A=}&(a_{11},a_{12},a_{22},a_{23},a_{33},a_{34},a_{44}) \\ 
{\bf IA=}&(1,3,5,7,8) \\
{\bf JA=}&(1,2,2,3,3,4,4) 
\end{array}
$$

\end{description}




\subsection{Search contour}

\begin{figure}[htbp]
\begin{small}
 \center{\includegraphics[width=0.95\textwidth]
        {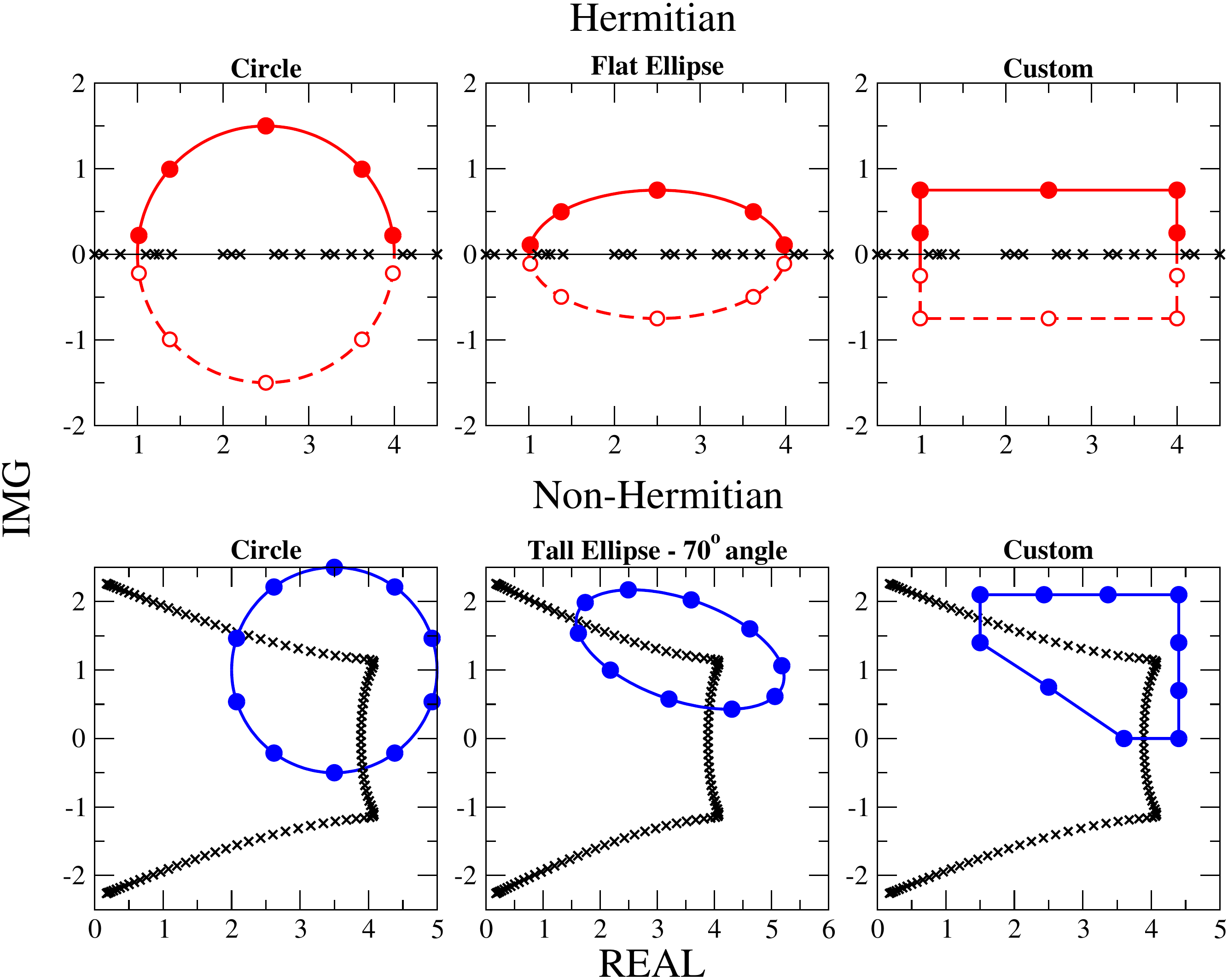}}
        \caption{\em \label{fig_contour} Various search contour examples for the Hermitian and the non-Hermitian/Polynomial FEAST algorithms.
Both algorithms feature standard ellipsoid contour options and the possibility to define custom arbitrary shapes. 
In the Hermitian case, the contour is symmetric with the real axis and only the nodes in the upper-half may be generated.
In the non-Hermitian/Polynomial case, a full contour is needed to enclose the wanted complex eigenvalues. Some data used to generate these plots:\\
{\bf Hermitian case:} {\tt fpm(2)=5} for all, $[E_{min}, E_{max}]=[1,4]$, $r=1.5$ for all;   {\tt fpm(18)=50} for the flat ellipse; and expert routine for
the custom contour \\
{\bf Non-Hermitian/Polynomial case:} {\tt fpm(8)=10} for all; $E_{mid}=3.5+i$ and $r=1.5$ for circle; $E_{mid}=3.4+1.3i$, $r=0.75$,
 {\tt fpm(18)=200}, {\tt fpm(19)=70} for tall rotated ellipse; and expert routine for
the custom contour 
}
\end{small}
\end{figure}

Figure \ref{fig_contour} summarizes the different search contour options possible 
for both the Hermitian and non-Hermitian (including Polynomial) FEAST algorithms.

For the Hermitian case, the user must then specify a 1-dimensional real-valued search interval $[E_{min}, E_{max}]$. 
These two points are used to define a circular or ellipsoid contour $\cal C$ centered on the real axis, and along
 which the complex integration nodes are generated. 
The choice of a particular quadrature rule 
will lead to a different set of relative positions for the nodes and associated quadrature weights. 
Since the eigenvalues are real, it is convenient to select a symmetric contour 
with the real axis (${\cal C}= {\cal C}^*$) since it only requires to operate the quadrature
on the half-contour (e.g. upper half). 

With a non-Hermitian/Polynomial problem, it is necessary to specify  a 2-dimensional search interval that surrounds
the wanted complex eigenvalues. 
Circular or ellipsoid contours can also be used and they can be generated using standard options included into FEAST v4.0. 
These are defined by a complex midpoint $E_{mid}$ and a radius $r$ for a circle (for an ellipse the ratio 
between the horizontal axis $2r$ and vertical axis can also be specified, as well as an angle of rotation). 
A ``Custom Contour'' feature is also supported that can  use  
arbitrary  quadrature nodes and weights (provided by the users).

The FEAST package provides a couple of utility routines that 
can return the integration nodes and weight used by the FEAST interfaces:
\begin{description}
\item[{\tt zfeast\_contour(Emin,Emax,fpm2,fpm16,fpm18,Zne,Wne)}] ~\\  Returns FEAST integration nodes and weights for a half-contour contour defined by {\tt Emin} and {\tt Emax}. To be used with Hermitian FEAST interfaces.

\item[{\tt zfeast\_gcontour(Emid,r,fpm8,fpm16,fpm18,fpm19,Zne,Wne)}]~\\ -  Returns FEAST integration nodes and weights for a full contour defined by {\tt Emid} and {\tt r}. To be used with non-Hermitian and polynomial FEAST interfaces.
\end{description}

The description of the arguments list for these routines is given in Table~\ref{tab_contour} and Table~\ref{tab_gcontour}.

\begin{table}[htbp]
\begin{center}
\begin{small}
\begin{tabular}{|l|l|c|l|}
\hline
 & Type & I/O &Description \\ 
\hline  \hline
{\tt Emin, Emax}  & 
double real  & in & Lower and Upper bounds of search interval \\ \hline
{\tt fpm2}  & 
integer  & in & Value of {\tt fpm(2)}- \#contour point (half-contour)\\ \hline
{\tt fpm16}  & 
integer  & in & Value of {\tt fpm(16)}- Integration type\\ \hline
{\tt fpm18}  & 
integer  & in & Value of {\tt fpm(18)}- Ellipse definition\\ \hline
{\tt Zne}  &   double complex    &  out  & Integration nodes\\  
{\tt Wne}  &   double complex      &  out  & Integration weights\\   \hline
\end{tabular}
\caption{\label{tab_contour} List of arguments for {\tt zfeast\_contour}.}
\end{small}
\end{center}
\end{table}

\begin{table}[htbp]
\begin{center}
\begin{small}
\begin{tabular}{|l|l|c|l|}
\hline
 & Type & I/O &Description \\ 
\hline  \hline
{\tt Emid}  & 
double complex  & in & Coordinate center of the contour ellipse\\ \hline
{\tt r}  & 
double real  & in & Horizontal radius of the contour ellipse\\ \hline
{\tt fpm8}  & 
integer  & in & Value of {\tt fpm(8)}- \#contour point (full-contour)\\ \hline
{\tt fpm16}  & 
integer  & in & Value of {\tt fpm(16)}- Integration type\\ \hline
{\tt fpm18}  & 
integer  & in & Value of {\tt fpm(18)}- Ellipse definition\\ \hline
{\tt fpm19}  & 
integer  & in & Value of {\tt fpm(19)}- Ellipse rotation angle\\ \hline
{\tt Zne}  &   double complex    &  out  & Integration nodes\\  
{\tt Wne}  &   double complex      &  out  & Integration weights\\   \hline
\end{tabular}
\caption{\label{tab_gcontour} List of arguments for {\tt zfeast\_gcontour}.}
\end{small}
\end{center}
\end{table}

\subsection{Contour Customization}

The Custom Contour feature grants the flexibility to target specific eigenvalues in a complex plane. This feature 
must be used with ``Expert'' routines that take two additional arguments containing the complex integration nodes and weights. Custom contours can be employed by following three simple steps: 
\begin{enumerate}
\itemsep 1pt
\parskip 1pt
\item Define a contour (half-contour that encloses [$\lambda _{min}$, $\lambda _{max}$] for the Hermitian problem, or full
contour for the non-Hermitian/Polynomial problem), 
\item Calculate corresponding integration nodes and weights, and 
\item Call ``Expert'' FEAST routine (either Driver or RCI interfaces by adding a {\tt x} at the end of the routine name).
\end{enumerate}

Furthermore, the FEAST package provides a utility  routine {\tt zfeast\_customcontour} that can assist the user to extract 
nodes and weights from a custom design arbitrary geometry in the complex plane (full-contour).
Users must only define the geometry of their contour. 
The contour can be comprised of line segments and half ellipses. 
Two important points to note: (i) the actual contour will end up being  a polygon defined by the integration points 
along the path, and  (ii) only convex contours may be used. A geometry that contains {\tt P} contour parts/pieces 
is defined using three arrays {\tt Zedge}, {\tt Tedge}, and {\tt Nedge}. The interface is defined below and
the description of the arguments list is given in Table~\ref{tab_custom}. 

\begin{quote}
\begin{tt}
zfeast\_customcontour(Nc,P,Nedge,Tedge,Zedge,Zne,Wne)
\end{tt}
\end{quote}

\begin{table}[htbp]
\begin{center}
\begin{small}
\begin{tabular}{|l|l|c|l|}
\hline
 & Type & I/O &Description \\ 
\hline  \hline
{\tt Nc}  & 
integer  & in & The total number of integration nodes, should be equal to \\
   &      &    & {\tt SUM(Nedge(1:P))} \\ \hline
{\tt P}  & 
integer  & in & Number of contour parts/pieces that make up the contour\\ \hline
{\tt Zedge} & integer({\tt P}) & in & Complex endpoints of each contour piece \\
         &                     &   & Remark: * endpoints positioned in clockwise direction \\
         &                     &   &~~~~~~~~~~~~~* the $k^{th}$ piece is [{\tt Zedge($k$),Zedge($k+1$)}] \\  
         &                     &   &~~~~~~~~~~~~~* last piece is [{\tt Zedge(P),Zedge(1)}]  \\ \hline
{\tt Tedge} & integer({\tt P}) & in &   The type of each contour piece: \\ 
            &                  &    &   *If {\tt Tedge($k$)}=0,   $k^{th}$ piece is a line \\
            &                  &    &   *If {\tt Tedge($k$)}$>$0, $k^{th}$ piece is a (convex) half-ellipse \\
            &                  &    &   ~~  with {\tt Tedge(k)}/100 = ratio $a$/$b$ and $a$ primary radius from the endpoints \\
            &                  &    &   ~~~Remark: 100 is a half-circle \\ \hline
{\tt Nedge} & integer({\tt P}) & in & \#integration intervals to consider for each piece \\
            &                  &    & define the accuracy of the trapezoidal rule by piece for FEAST\\ \hline
{\tt Zne}  &   double complex    &  out  & Custom integration nodes for FEAST\\  
{\tt Wne}  &   double complex      &  out  & Custom integration weights for FEAST\\   \hline
\end{tabular}
\caption{\label{tab_custom} List of arguments for {\tt zfeast\_customcontour}.}
\end{small}
\end{center}
\end{table}

As an example, the following code will generate the corresponding complex contour.

\noindent
\begin{minipage}{0.6\textwidth}
\begin{footnotesize}
\begin{F90}
P = 3 ! number of pieces that make up the contour 
allocate( Zedge(1:P),Nedge(1:P),Tedge(1:P) )
Zedge = (/(0.0d0,0.0d0),(0.0d0,1.0d0),(6.0d0,1.0d0)/)
Tedge(:) = (/0,0,50/)! (line)--(line)--(half-circle)
Nedge(:) = (/8,8,8/) ! integration intervals by piece
Nc = sum(Nedge(1:P)) ! #contour points (here 24)
allocate( Zne(1:Nc), Wne(1:Nc) ) 
call zfeast_customcontour(Nc,P,Nedge,Tedge,Zedge,Zne,Wne)
! (Zne, Wne) are now defined and ready to use for FEAST
\end{F90}	
\end{footnotesize} 
\end{minipage}
\hfill
\begin{minipage}{0.39\textwidth}
\centering
\includegraphics[width=\textwidth,angle=0]{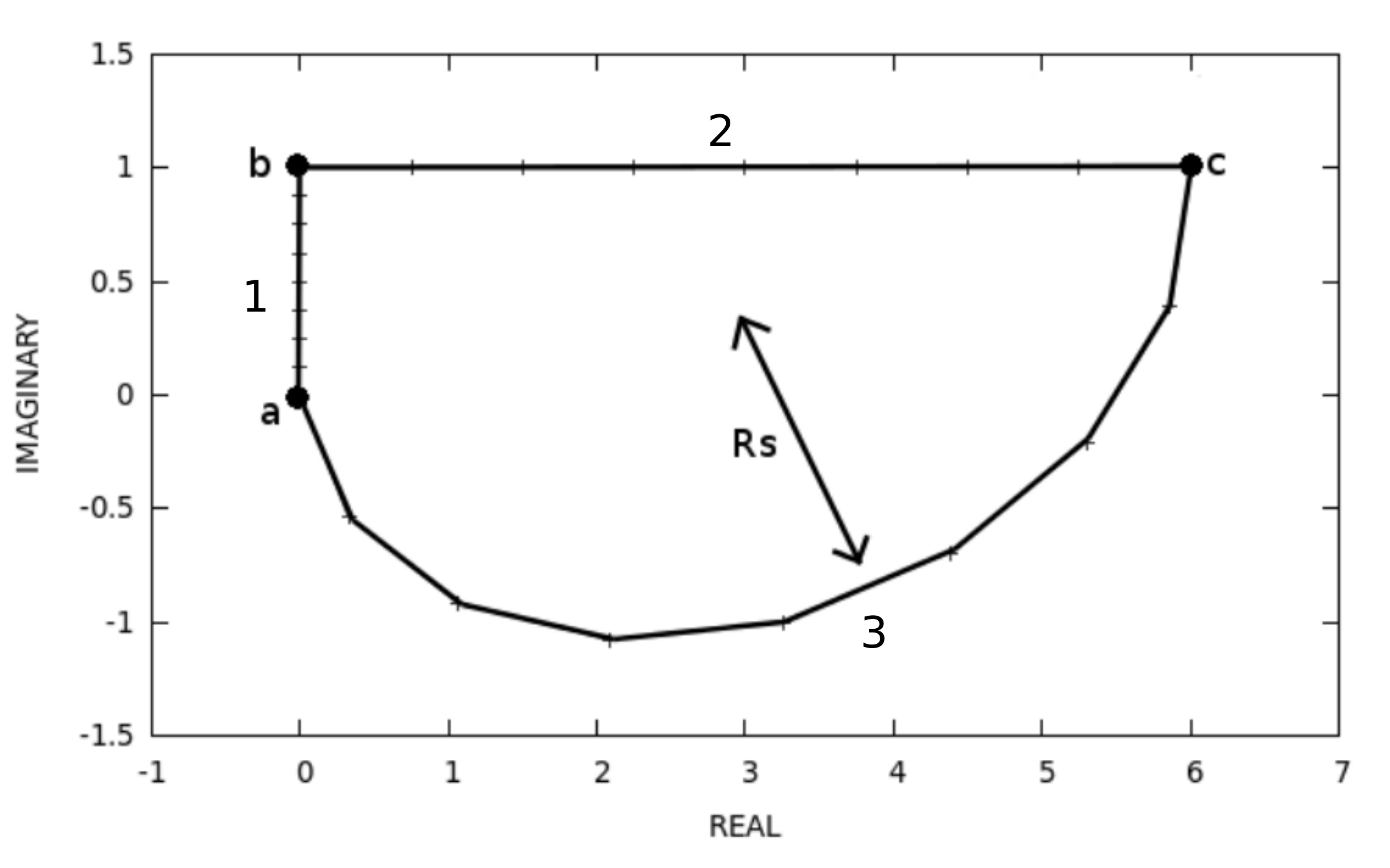}
\end{minipage}

The {\tt \$FEASTROOT/example/FEAST} directory provides Fortran and C implementation of the expert FEAST routines using a custom contour. It is applied on the System4 example using both dense, banded and sparse-CSR storage. Here, the complete list of routines:

\begin{center}
\begin{tabular}{ll}
&  \multicolumn{1}{c}{System4} \\ \hline\hline
\col dense  & \col \\
& {\tt \{F90,C\}dense\_zfeast\_syevx} \\
\col banded  & \col \\
& {\tt \{F90,C\}dense\_zfeast\_sbevx} \\
\col sparse  & \col \\
  & {\tt \{F90,C\}dense\_zfeast\_scsrevx} \\
\hline
\end{tabular}
\end{center}

\newpage



\subsection{FEAST utility sparse drivers}
If a sparse matrix can be provided by the user in coordinate/matrix market format,
the {\tt \$FEASTROOT/utility} directory offers a quick way  
to test all the FEAST parameter options and 
the efficiency/reliability/timing of the FEAST SPARSE driver interfaces.
Two general drivers are provided for FEAST/IFEAST and PFEAST/PIFEAST,  named {\tt driver\_feast\_sparse}
or  {\tt driver\_pfeast\_sparse} in their respective subdirectories. The command ``{\tt>make all}'' should compile the drivers.

If we denote {\tt mytest} a generic name for the user's eigenvalue system test $\bf Ax=\lambda x$
or  $\bf Ax=\lambda Bx$. You will need to create the following three files:
     \begin{itemize}
       \item {\tt mytest.mtx} should contain the matrix {\bf A} in coordinate format; As a reminder,
         the coordinate format is defined row by row as
         
\begin{shaded}
\begin{footnotesize}
\begin{Verbatim}[frame=single,baselinestretch=0.9]
      N       N       NNz
      :       :       :              :
      i       j      real(valj)    img(valj)
      :       :       :              : 
      :       :       :              :
      :       :       :              :
     iNNZ    jNNZ    real(valNNZ) img(valNNZ)
\end{Verbatim}
\end{footnotesize}
\end{shaded}
with {\tt N}: size of matrix, and {\tt NNZ:} number of non-zero elements.
       \item {\tt mytestB.mtx} should contain the matrix {\bf  B} (if any) in coordinate format;
       \item {\tt mytest.in} should contain the search interval, selected FEAST parameters, etc. 
The following {\tt .in} file is given as a template example (here for solving a standard eigenvalue problem in 
double precision):
    
\begin{shaded}
\begin{footnotesize}
\begin{Verbatim}[frame=single,baselinestretch=0.9]
s       ! s: symmetric, h: hermitian, g: general
g       ! e=standard or g=generalized eigenvalue problem
d       ! (d,z) precision i.e (double real, double complex)
F       ! UPLO (L: lower, U: upper, F: full) for the coordinate format of matrices
0.18d0  ! Emin
1.00d0  ! Emax
25      ! M0 search subspace (M0>=M)
2       !!!!!!!!!! How many changes from default fpm(1,64) (use 1-64 indexing)
1 1     !fpm(1)=1 !example comments on/off  (0,1)
2 4     !fpm(2)=4 !number of contour points
\end{Verbatim}
\end{footnotesize}
\end{shaded}
You may change any of the above options to fit your needs. For example, you could add as many fpm FEAST parameters as you wish.
You can also use the flag {\tt fpm(43)=1} to switch to IFEAST.
In addition, the {\tt L} or {\tt U} options for {\tt UPLO}  give you the possibility
to provide only the lower or upper triangular part of the matrices {\tt mytest.mtx} and {\tt mytestB.mtx} 
 in coordinate format.
\end{itemize}

Finally results and timing can be obtained by running the FEAST sparse driver: 
\begin{quote}
\fbox{\tt ./driver\_feast\_sparse <PATH\_TO\_MYTEST>/mytest} 
\end{quote}
In addition, all the eigenvalue solutions will be locally saved in the file {\tt eig.out}.\\

For the 
PFEAST sparse drivers, a run would look like (other options could be applied):

 \fbox{\parbox{\dimexpr\linewidth-2\fboxsep-2\fboxrule\relax}{\tt mpirun  -env~ MKL\_NUM\_THREADS~ <omp> -ppn~ 1 -n <nL1> ./driver\_pfeast\_sparse \\
  {\color{white}{aaaaaaaaaa}}<PATH\_TO\_MYTEST>/mytest nL3}}

where ${\tt <nL1>}$ represents the total number of MPI processes to use for a single contour, and {\tt nL3} is the number
of MPI processes used for solving the linear systems.
As a reminder, L3 uses MKL-Cluster-PARDISO with PFEAST, and PBicGStab with PIFEAST. The level L3 can also be threaded by  setting the
shell variable  {\tt MKL\_NUM\_THREADS} equal to the desired number of threads. Make sure that {\tt <omp>}*{\tt <nL1>} does not exceed
the number of physical cores. Several  combinations of {\tt <nL1>}, {\tt <nL3>} and {\tt <omp>} are
possible depending also on the value of the {\tt -ppn} directive.\\

In order to illustrate a direct use of the utility drivers,  several examples are provided 
in the directory {\tt \$FEASTROOT/utility/data} summarized in Table~\ref{tab_data}.

\begin{table}[h!]
\begin{small}
\begin{center}
\begin{tabular}{|l|cc||ccc||cc|}
  \multicolumn{1}{c}{}            &  \multicolumn{1}{c}{Real}   &   \multicolumn{1}{c}{Complex}  &
  \multicolumn{1}{c}{Symmetric} &   \multicolumn{1}{c}{Hermitian} &   \multicolumn{1}{c}{General} &
  \multicolumn{1}{c}{Standard} &   \multicolumn{1}{c}{Generalized} \\ \hline\hline
helloworld      &  \col X   &            &        \col X   &            &           &       \col  X  &            \\ \hline
system1         & \col  X     &            &   \col   X     &            &           &            &       \col X      \\ \hline
system2         &        &   \col  X      &            &    \col   X    &           &   \col    X    &             \\ \hline
system3         &  \col X     &            &            &            &    \col   X   &            &  \col     X     \\ \hline
system4         &        &  \col   X      &  \col    X     &            &           &   \col    X    &             \\ \hline
cnt             & \col  X     &            &   \col   X     &            &           &            &   \col    X     \\ \hline
co              &  \col X     &            &    \col  X     &            &           &            &   \col    X     \\ \hline
c6h6            &  \col X     &            &    \col  X     &            &           &            &   \col    X      \\ \hline
Na5             &  \col X     &            &    \col  X     &            &           &     \col  X    &             \\ \hline
grcar           &  \col X     &            &            &            &     \col  X   &     \col  X    &             \\ \hline
qc324           &        &  \col   X      &   \col   X     &            &           &    \col   X    &              \\ \hline
bcsstk11        &  \col X     &            &    \col  X     &            &           &            &   \col    X       \\ \hline
\end{tabular}
\end{center}
\caption{\label{tab_data} List of system matrices provided in the {\tt \$FEASTROOT/utility/data} directory. System 1 to 4 corresponds
to the matrices used in the {\tt example} directory.}
\end{small}
\end{table}

To run a specific test, you can execute using the FEAST driver (for example): 
\begin{quote}
\fbox{\tt ./driver\_feast\_sparse ../data/cnt} 
\end{quote}

or using the PFEAST driver (for example):
\begin{quote}
\fbox{\tt mpirun -env MKL\_NUM\_THREADS 2 -n 4 ./driver\_pfeast\_sparse ../data/cnt 2} 
\end{quote}

Here we use 8 total compute cores and  {\tt nL1=4} MPI, {\tt nL2=nL1/nL3=2} MPI, {\tt nL3=2} MPI, {\tt omp=2} threads.





\end{document}